\documentclass[aip,reprint]{revtex4-1}

\usepackage{graphicx}
\usepackage{dcolumn}
\usepackage{bm}
\usepackage{hyperref}
\usepackage[utf8]{inputenc}
\usepackage[T1]{fontenc}
\usepackage{mathptmx}
\usepackage{etoolbox}
\usepackage{xcolor}
\newcommand{\tg}[1]{\textcolor{black}{#1}} 
\newcommand{\jb}[1]{\textcolor{black}{#1}}
\draft 
\usepackage{float}
\begin{document}


\title{Attractive carbon black dispersions: structural and mechanical responses to shear} 



\author{Julien Bauland}
\affiliation{Univ Lyon, Ens de Lyon, CNRS, Laboratoire de Physique, 69342 Lyon, France}

\author{Louis-Vincent Bouthier}
\affiliation{Groupe CFL, CEMEF, Mines Paris PSL, 1 Rue Claude Daunesse, 06904 Sophia Antipolis,France}



\author{Arnaud Poulesquen}
\affiliation{CEA, DES, ISEC, DE2D, SEAD, LCBC, Université of Montpellier, Marcoule, France}

\author{Thomas Gibaud}
\email[]{Corresponding author, thomas.gibaud@ens-lyon.fr}
\affiliation{Univ Lyon, Ens de Lyon, CNRS, Laboratoire de Physique, 69342 Lyon, France}

\date{\today}
\begin{abstract}
The rheological behavior of colloidal dispersions is of paramount importance in a wide range of applications, including construction materials, energy storage systems and food industry products. These dispersions consistently exhibit non-Newtonian behaviors, a consequence of intricate interplays involving colloids morphology, volume fraction, and inter-particle forces. Understanding how colloids structure under flow remains a challenge, particularly in the presence of attractive forces leading to clusters formation. In this study, we adopt a synergistic approach, combining rheology with ultra small-angle X-ray scattering (USAXS), to probe the flow-induced structural transformations of attractive carbon black (CB) dispersions and their effects on the viscosity. Our key findings can be summarized as follow.
First, testing different CB volume fractions, in the high shear rate hydrodynamic regime, CB particles aggregate to form fractal clusters. Their size conforms to a power law of the shear rate, $\xi_c \propto \dot{\gamma}^{-m}$, with $m\simeq 0.5$.
Second, drawing insights from the fractal structure of clusters, we compute an effective volume fraction $\phi_{\mathrm{eff}}$ and find that microstructural models adeptly account for the hydrodynamic stress contributions. We identify a critical shear rate $\dot{\gamma^*}$ and a critical volume fraction $\phi_{\mathrm{eff}}^{*}$, at which the clusters percolate to form a dynamical network. 
Third, we show that the apparent yield stress measured at low shear rates inherits its properties from the percolation point.
Finally, through data scaling and the integration of the Einstein's viscosity equation, we revisit and discuss the Caggioni-Trappe-Spicer model, revealing a significant connection between its empirical parameters and the structural properties of CB dispersions under flow.
\end{abstract}


\maketitle 


\section{Introduction}

Colloidal dispersions are encountered in numerous application fields, for instance as cement paste in construction industry~\cite{Chougnet2008}, as drilling muds during extraction operations~\cite{Morariu2022}, as flowable electrodes used for energy storage~\cite{Alfonso2021}, or as oil in water emulsions in food industry ~\cite{Genovese2007,Xi2019}. In the aforementioned examples, predicting and controlling the viscosity of the dispersions is of key importance to optimize their use. The rheological behavior of colloidal dispersions is usually non-Newtonian, and depends on the shape, volume fraction and interaction potential between the particles~\cite{Mewis2011}. In scenario with low volume fractions $\phi_{r_0}$ and strong attractive forces, colloidal dispersions become unstable, and particles of radius $r_0$ aggregate into "flocs" or "clusters" of radius $\xi_c$. These clusters exhibit a fractal structure and the number of particles $N_c$ in a cluster scales as~\cite{Lazzari2016}:

\begin{equation}
    N_c = \bigg(\frac{\xi_c}{r_0}\bigg)^{d_f}
    \label{eq:fractalcluster}
 \end{equation}
 
\noindent with $d_f$ the fractal dimension of clusters. \jb{Because of their fractal nature, clusters cannot continue growing indefinitely as the volume they occupy increases faster than the available space. Therefore, there may come a point where the spheres enclosing the clusters reach a random close packing limit. At this limit, the spatial separation between the clusters approaches zero, allowing them to interconnect through the outermost particles on their respective surfaces~\cite{zaccone2014}. This outcome is valid at rest~\cite{haw1995} and under shear~\cite{xie2010}}. The shear-dependent structure of clusters is a crucial factor for understanding the dispersion properties, whether the dispersion is at rest or subjected to shear. 

Under quiescent conditions, aggregation is driven by the thermal energy. The aggregation rate is either limited by particles diffusion~\cite{weitz1984} or by the reaction probability upon collision~\cite{schaefer1984}, leading to loosely structured clusters~\cite{Lazzari2016} with $1.7 < d_f < 2.1$. When the volume fraction is sufficiently high, further aggregation of clusters leads to the formation of a space-spanning network called a "gel", that displays an elastic modulus and a yield stress~\cite{trappe2004}. 

When attractive dispersions experience flow, clusters tend to condense, resulting in $d_f > 2.4$~\cite{Sonntag1986,Zaccone2009,Massaro2020}. The equilibrium cluster size $\xi_c$ is determined by the balance between the viscous drag force acting on a particle ($F_\mathrm{visc}$) and the attractive forces holding particles together ($F_\mathrm{attr}$). This balance is quantified by the Mason number, expressed as~\cite{Mewis2009}:
\begin{equation}
    \mathrm{Mn} = \frac{F_\mathrm{visc}}{F_\mathrm{attr}}=\frac{6\pi \eta r_0^2 \dot{\gamma}}{U/\delta}
    \label{eq:mason}
 \end{equation}
\noindent with $r_0$ (m) the particle radius,  $\dot{\gamma}$ the shear rate and $\eta$ a viscosity that we will define later. $U$ (N.m) and $\delta$ (m) are the depth and width of the attractive interaction potential, respectively. An important question revolves around the nature of the viscous force involved in Eq.~\ref{eq:mason} and the stress experienced by clusters. From the dependence of the cluster size with the shear rate, it was argued that clusters are sufficiently distant, and that the fluid stress $\sigma_f = \eta_f\dot{\gamma}$ was the primary factor~\cite{Sonntag1986,Varga2018} with $\eta_f$ the viscosity of the background fluid. An alternative proposition suggests that interactions between clusters would lead to a substantial increase in hydrodynamic stress~\cite{Hipp2021} and that the dispersion viscosity $\eta$ that must be taken into account in Eq.~\ref{eq:mason} to explain the equilibrium cluster size. This debate underscores the complex interplay of factors involved in cluster formation and behavior in attractive colloidal dispersions submitted to flow. The choice between the fluid stress or the dispersion stress as the key determinant might depends on the characteristics of the dispersion system and the experimental conditions. In all cases, the Mason number was effectively reported to control the cluster size at high shear~\cite{Bouthier2022a,Hipp2021,Varga2018}.

Under this framework, the structuring degree of the dispersion is set by the flow, and in return the flow is affected by the structure of the dispersion. These reversible and time-dependent flow-induced structural changes are called "thixotropy"~\cite{Mewis2009}. Macroscopically, attractive dispersions usually show two regimes, depending on the $\mathrm{Mn}$ value. 
At high $\mathrm{Mn}$, the shear rate dependence of the cluster size leads to a shear thinning behavior and the clusters size was reported to scale as
\begin{equation}
   \xi_c \propto \mathrm{Mn}^{-m},
   \label{eq:cluster}
 \end{equation}
where $m$ is the breaking exponent. There are yet no consensus on the value of the breaking exponent $m$. Wessel and Ball~\cite{Wessel1992} first proposed that the maximum cluster size should scale with the hydrodynamic stress with $m=1/3$. More complex models distinguish highly deformable clusters from rigid clusters, yielding $m = 1/2$ and $m=1/3$, respectively~\cite{Snabre1996a, potanin1993}. Recently, Bouthier et al.~\cite{Bouthier2022a} proposed that $m$ should vary with the fractal dimension of clusters, with $m=1/(1+d_f)$. Besides analytical calculations, experiments~\cite{Hipp2021,Sonntag1986,hunter1980,brakalov1987} and simulation~\cite{Varga2018, ruan2020} results have reported values for $m$ ranging from 0.2 to 1.  At low values of $\mathrm{Mn}$, the cluster size becomes independent of the shear rate~\cite{Varga2018} as clusters fill the space, forming a transient network and leading to an apparent yield stress $\sigma_y$. Varga and Swan~\cite{Varga2018} proposed that percolation would occur at a critical value of the Mason number, $\mathrm{Mn}_c \propto \phi_{r_0}^{2/(3-d_f)}$ with $\phi_{r_0}$ the volume fraction of particles.

In practice, the rheological properties of attractive dispersions are assessed by applying a given shear rate $\dot{\gamma}$ and measuring the resulting stress with $\sigma=\eta\dot{\gamma}$. To assess the dependence of the dispersions viscosity with the shear rate, a flow curve test is commonly conducted. This test consists in applying a stepwise increment in shear rate with a given step duration $\Delta t$. Due to the thixotropic nature of attractive dispersions, a finite amount of time is necessary to reach the equilibrium viscosity. Consequently, a rapid increase in shear rate will only characterize the transient viscosity. Flow curves, whether represented as $\sigma$ $vs$ $\dot{\gamma}$ or $\eta$ $vs$ $\dot{\gamma}$, are usually described using the Herschel-Bulkley model~\cite{herschel1926}:  
\begin{equation}
    \sigma = \sigma_y \bigg[ 1 + \left(\frac{\dot{\gamma}}{\dot{\gamma}_{HB}^*}\right)^{n_{HB}} \bigg], \\ 
        \label{eq:hb}
\end{equation}

\noindent where $\sigma_y$ is the yield stress (Pa), $\dot{\gamma}_{HB}^*$ a critical shear rate and $n_{HB}$ an empirical exponent. This empirical model effectively captures both the yield stress and the shear-thinning behavior of attractive dispersions with $n_{HB} < 1$. However, it does not provides any microscopic interpretation of its parameters as it dismiss the cluster properties. 
\tg{In more developed models, like structural kinetics model~\cite{Mewis2009}, the rheological parameters of the constitutive equation are set as a function of a structural parameter $\lambda$, comprised between 0 and 1 for a totally broken or fully developed cluster structure, respectively. By introducing kinetics equation for $\lambda$, the structural kinetics models efficiently capture the time-dependent properties of dispersions, but the parameter $\lambda$ cannot be easily related with any physical aspects of the microstructure. In population balanced models~\cite{jeldres2018}, a partial differential equation tracks the evolution of the size distribution function with respect to time. The equation includes terms that represent the rates of various particle interactions or transformations. Solving this equation provides insights into the evolution of the distribution of particle sizes over time and under different flow conditions.}

\tg{In this study, we choose to focus on microstructural models, which directly rely on structural changes that can experimentally be measured}. In such models, the shear stress is decomposed as a sum of an elastic $\sigma_e$ and an hydrodynamic $\sigma_h$ contribution, so that $\sigma = \sigma_e +\sigma_h$. The elastic stress originates from the network and is typically treated as a constant, derived from the particle interactions and the cluster properties~\cite{DeRooij1994,Potanin1995}. The hydrodynamic stress $\sigma_h$ is calculated by introducing an effective volume fraction of clusters $\phi_{\mathrm{\mathrm{eff}}}$. $\phi_{\mathrm{\mathrm{eff}}}$ is the equivalent volume fraction of spheres, considering the fractal structure of clusters \cite{DeRooij1994,Potanin1995,Wolthers1996}:
\begin{equation}
    \phi_{\mathrm{eff}} = \frac{\phi_{r_0}}{k} \bigg(\frac{\xi_c}{r_0}\bigg)^{3-d_f} \\ 
        \label{eq:phieff}
\end{equation}

\noindent where $\phi_{r_0}$ is the volume fraction of primary particles and $k$ is a scaling constant usually approximated as unity. Then, various models have been used to relate $\phi_{\mathrm{eff}}$ with the viscosity of the dispersion $\eta$~\cite{quemada2002, silbert1999, Mewis2009, varadan2001}. Assuming that the viscosity dependence on the volume fraction follows the one of a hard sphere dispersion, a generalized viscosity equation for concentrated regimes can be used,~\cite{Quemada1978,Genovese2012} such as the one proposed by Krieger and Dougherty~\cite{Krieger1959}:

\begin{equation}
    \eta = \eta_f \bigg(1-\frac{\phi_{\mathrm{eff}}}{\phi_{\mathrm{eff}}^M}\bigg)^{-2.5\phi_{\mathrm{eff}}^M} 
            \label{eq:etaKD}
\end{equation}

\noindent with $\phi_{\mathrm{eff}}^M$ the maximum packing fraction of hard spheres, typically $\phi_{\mathrm{eff}}^M\simeq 0.64$. 

In summary, a significant body of work aimed to establish connections between the cluster properties and the viscosity of colloidal dispersions. However, few studies have probed both aspects simultaneously and structural characterizations are lacking. Consequently, the validity of microstructural models has not been thoroughly tested~\cite{Mewis2009} and some questions have still to be answered: How the cluster properties change with shear rate?  Are cluster properties driven by stress or shear rate? Is the decomposition of the stress into an elastic and hydrodynamic contribution sufficient to model a flow curve?

To address these questions, we follow an experimental approach with carbon black (CB) dispersions in oil, chosen as study model. Such dispersions present peculiar and fundamental rheological properties~\cite{Gibaud2020b, Richards2023,Trappe2000} including rheopexy~\cite{Ovarlez:2013,Helal:2016,Hipp2019}, delayed yielding~\cite{Gibaud:2010,Grenard:2014}, fatigue~\cite{Gibaud:2016, Perge:2014}, rheo-conductive properties~\cite{Richards2017}, sensitivity to flow cessations~\cite{Dages2022b, bouthier2023} and rheo-acoustic properties~\cite{Gibaud2020a, Dages2021}. Furthermore, CB dispersions directly find utility in numerous applications such as ink~\cite{liu2021}, cement~\cite{li2006} or semi-solid flow batteries~\cite{liu2023}. 


The paper is structured as follows. We initially characterize the flow properties of CB dispersions using rheology across a range of volume fractions spanning from 0.6\% to 4.1\%. Then, we examine the structural features of these dispersions under flow conditions using ultra small-angle X-ray scattering (USAXS)~\cite{Panine2003}. Finally, we establish a relationship between the shear-dependent viscosity of the dispersions and their structural parameters by employing microstructural models.

\section{Material and methods}
\label{s:mm}
\subsection{Carbon black dispersions}

CB particles (Vulcan\textsuperscript{\textregistered}PF, Cabot, density $d_{cb} = 2.26 \pm 0.03$) were dispersed in mineral oil (RTM17 Mineral Oil Rotational Viscometer Standard, Paragon Scientific, viscosity $\eta_f = 252.1$ mPa.s at $T = 25^{\circ}$C, density $d_{oil} = 0.871$) at mass fractions $c_w$ ranging from 1.5 to 10 $\%$ ($w\backslash w$)~\cite{Dages2022b}. The corresponding volume fractions of CB particles $\phi_{r_0}$ were calculated as $ \phi_{r_0} = c_w/\bigg(c_w + \frac{d_{cb}}{d_{oil}}(1-c_w)\bigg) $. After mixing, dispersions were sonicated during 2h in an ultrasonic bath (Ultrasonic cleaner, DK Sonic\textsuperscript{\textregistered}, United-Kingdom) to ensure that all particles were fully dispersed.
\subsection{Rheology}

Experiments were carried out with two stress controlled rheometers: ($i$) a Haake RS6000 (Thermo Scientific) and ($ii$) a MCR 301 (Anton Paar), both equipped with a Couette geometry composed of concentric polycarbonate cylinders (inner diameter 20~mm, outer diameter 22~mm and height 40~mm). 

All rheological measurements were performed at 25$^{\circ}$C. After loading the CB dispersion, a preshear, $\dot{\gamma} = 1000$ s$^{-1}$, was first applied during 60 s to erase any shear history. 
Then, a flow curve test was performed by ramping downward from $\dot{\gamma}=$1000 to 0.01 s$^{-1}$ (10 pts per decade) with a duration $\Delta t = 1$~s/pts and measuring the resulting stress $\sigma$. A downward ramping of the shear rate was chosen to prevent heterogeneous flows (\emph{e.g.} shear banding) usually observed for yield stress fluids in the vicinity of the yield stress \cite{Divoux2013}. We varied $\Delta t$ and probed 50 and 100 s/pts and found that the flow curves were identical down to $\dot{\gamma}\sim 1$~s$^{-1}$ (see Appendix, Fig.~\ref{fig:FCtime}). For shear rates $\dot{\gamma} > 1$~s$^{-1}$ a stationary regime is rapidly found as observed in~\cite{Dages2022b}. Below $\dot{\gamma}\sim 1$~s$^{-1}$, the rheology is not stationary and wall slip, sedimentation~\cite{Hipp2019} or shear banding~\cite{Gibaud:2010} may occur. 
It is noteworthy that Hipp et al.~\cite{Hipp2021} also performed flow curve tests on CB dispersions. However, they constructed "self-similar flow curves" through a series of rheological measurements that differ from a continuous ramping of the shear rate as performed here. Consequently, the shear history undergone by the CB dispersions before each shear rate is significantly different.


\begin{figure}
    \includegraphics[scale=0.55, clip=true, trim=0mm 0mm 0mm 0mm]{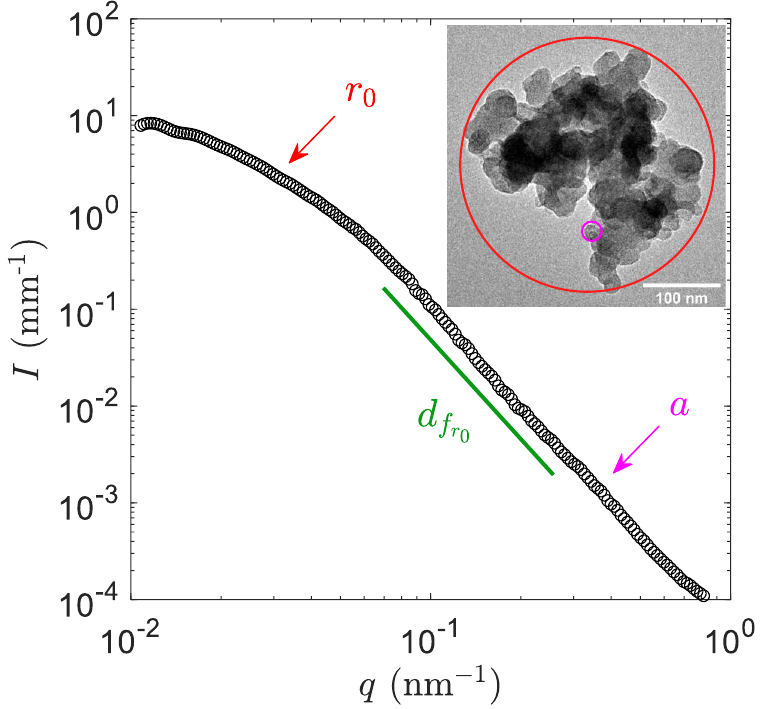}
    \centering
    \caption{Form factor of carbon black particles measured by USAXS for $\phi_{r_0}$ = 10$^{-4}$. Carbon black particles are composed of nodules of radius $a$ that are fused to form primary aggregates of radius $r_0$ and fractal dimension $d_{fr_0}$. Inset displays a representative electron microscopy image of a single CB particle. See Fig.~\ref{fig:FF} in Appendix for a detailed analysis of the form factor.}
    \label{figure0}
\end{figure}


\begin{figure*}
    \includegraphics[scale=0.5, clip=true, trim=0mm 0mm 0mm 0mm]{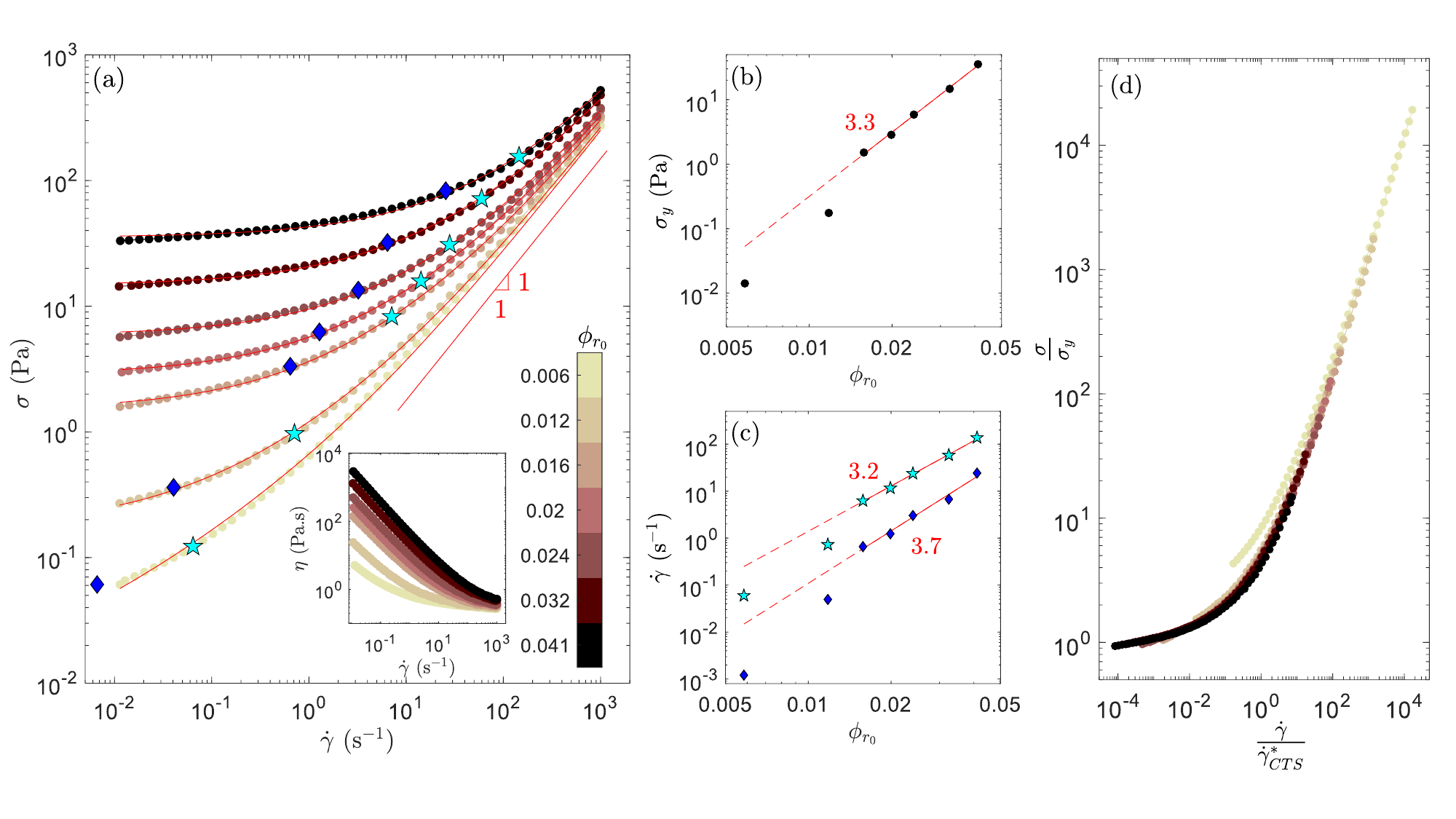}
    \centering
    \caption{Flow curve of carbon black (CB) dispersions and the Caggioni-Trappe-Spicer (CTS) model~\cite{Caggioni2020}. (a) Shear stress $\sigma$ $vs$ shear rate $\dot{\gamma}$ during fast flow curves of CB dispersions at various volume fractions $\phi_{r_0}$. The red curves correspond to the best fits with the CTS model (eq.~\ref{eq:trappe}). Inset: viscosity $\eta$ $vs$ $\dot{\gamma}$. Evolution of the parameters of the CTS model with $\phi_{r_0}$: (b) Apparent yield stress $\sigma_y$; (c) Critical shear rates $\dot{\gamma}_{CTS}^*$ (blue diamonds) and $\dot{\gamma}_{CTS}^p$ (cyan stars). $\dot{\gamma}_{CTS}^p$ is a "plastic" shear rate and $\dot{\gamma}_{CTS}^*$ is another critical shear rate obtained as $\dot{\gamma}_{CTS}^* = \sigma_y \slash \eta_0$, with $\eta_0$ the high shear viscosity. Red dash lines are the best power law fit of the data. (d) Normalized flow curves: $\sigma \slash \sigma_y$ $vs$ $\dot{\gamma} \slash \dot{\gamma}_{CTS}^*$.
    }
    \label{figure1}
\end{figure*}

\subsection{Ultra Small angle X-ray scattering (USAXS)}
The microstructural characteristics of the CB dispersion were probed using rheo-USAXS measurements conducted at the ID02 beamline within the European Synchrotron Radiation Facility (ESRF) in Grenoble, France~\cite{Narayanan2022}. The incident X-ray beam, with a wavelength of 0.1~nm, was collimated to dimensions of 50~$\mu$m vertically and 100~$\mu$m horizontally. Utilizing an Eiger2 4M pixel array detector, two-dimensional scattering patterns were acquired. The subsequent data reduction process is detailed in~\cite{Panine2003}. The scattering intensity $I(q)$ was derived by subtracting the two-dimensional scattering profiles of the carbon black dispersions and the mineral oil. Importantly, the resulting scattering intensity remained isotropic throughout this study (see Appendix, Fig.~\ref{fig:anisotropy}) and an azimuthal average was performed to obtain a one-dimensional $I(q)$. Measurements were conducted in both radial and tangential configurations, yielding equivalent results due to the isotropic nature of the dispersions microstructure over the tested $q$-range.

\subsection{Transmission electron microscopy}

A diluted dispersion of CB in absolute ethanol ($c_w \approx 2.10^{-5}$) was prepared and 5~\textmu L were deposited on a carbon grid (Carbon Film CF300-Cu, Electron Microscopy Science, England) and left to dry in a dust free environment. Individual CB particles were then imaged by transmission electron microscopy (TEM) using a JEOL JEM 1400 microscope. A representative set of CB particle images is displayed in Fig.~\ref{fig:FF} in Appendix. 

\subsection{Characterization of carbon black particles}

The CB particles were analyzed using TEM and USAXS. As shown in the TEM images in inset of Fig.\ref{figure0}, CB particles are composed of nodules of radius $a$ that are fused to form primary aggregates of radius $r_0$~\cite{Dannenberg2000}. These two length scales, $q_{r_0} \approx 5.10^{-2}$ and $q_a \approx 4.10^{-1}$ nm$^{-1}$ are also identified on the form factor of the CB particles, measured on a diluted sample ($\phi_{r_0}=$ 10$^{-4}$) at rest (Fig.\ref{figure0}). The power law scaling $I(q) \propto q^{d_{fr_0}}$ at intermediate $q$ is related to the fractal dimension of the primary aggregates $d_{fr_0}$. Having tested different models to fit the USAXS data (see Fig.~\ref{fig:FF} in Appendix), we estimate $a\simeq 20$~nm, $r_0\simeq 85$~nm and $d_{fr_0}\simeq 2.8$. Vulcan PF particles are therefore small and compact aggregates in comparison with other CB particles previously reported (e.g. $r_0 \simeq 180$~nm for Vulcan XC-72 \cite{Richards2017, Dages2021}). \tg{We note that the flow does not impact the integrity of CB particles, as evidence in Fig.~\ref{figure3}a-b, where $I(q)$ remains constant for $5.10^{-2}<q<2.10^{-1}$~nm$^{-1}$}. 


\section{Results and discussion}
\label{s:results}

\subsection{Rheological properties of carbon black dispersions}

We first examine the influence of the particles volume fraction $\phi_{r_0}$ on the flow behavior of the CB dispersions. Fig. \ref{figure1}a displays the flow curves for dispersions with $\phi_{r_0}$ ranging from 0.6 to 4.1\%. In Fig.~\ref{figure1}a, the flow curve exhibits a similar trend for all the tested volume fractions: a shear-thinning behavior at high shear rates and the presence of a dynamic yield stress $\sigma_y$ at low shear rates. The dynamic yield stress $\sigma_y$ increases with $\phi_{r_0}$ (Fig.~\ref{figure1}b) and its existence can even be debated at the lowest volume fraction ($\phi_{r_0}=0.006$). Dynamic oscillatory shear experiments indicate that the minimum particle volume fraction required to form a gel is $\phi_{r_0}^{gel} \approx 0.01$, a value consistent across different pre-shear histories (data not presented).

To quantitatively assess the impact of $\phi_{r_0}$ on the flow curves, two empirical models were employed: the Herschel-Bulkley~\cite{herschel1926} model and the Caggioni-Trappe-Spicer model~\cite{Caggioni2020}. The Herschel-Bulkley model fit is not entirely satisfactory, especially for $\dot{\gamma}<1~s^{-1}$ (see Appendix, Fig.~\ref{fig:HB}). Improved fits are obtained with the Caggioni-Trappe-Spicer model (CTS)~\cite{Caggioni2020}:
\begin{equation}
    \sigma = \sigma_y + \sigma_y\left(\frac{\dot{\gamma}}{\dot{\gamma}_{CTS}^p}\right)^{\frac{1}{2}} + \eta_0\dot{\gamma} \\ 
    \label{eq:trappe}
\end{equation}

\noindent where $\sigma_y$ is the yield stress (Pa), $\dot{\gamma}_{CTS}^p$ a critical shear rate (s$^{-1}$) and $\eta_0$ a viscosity (Pa.s). In addition to taking into account the elastic component represented by the yield stress $\sigma_y$, this model incorporates a Newtonian viscous stress at high shear rates, $\eta_0\dot{\gamma}$. Furthermore, it introduces a "plastic" dissipation mechanism, where the stress scales with the square root of the shear rate, serving as an intermediate regime between the purely viscous and purely elastic behaviors. The elastic and plastic regimes are separated by a critical shear rate $\dot{\gamma}_{CTS}^p$. A second critical shear rates $\dot{\gamma}_{CTS}^*$ can be obtained as $\dot{\gamma}_{CTS}^* =\sigma_y/\eta_0$. It corresponds to the point at which the stress contribution of the high shear viscosity equals the yield stress.~\cite{Caggioni2020} In Eq.~\ref{eq:trappe}, the high shear viscosity $\eta_0$ can be calculated with the Einstein's viscosity~\cite{einstein1905} $\eta_0 = \eta_{f}  (1+2.5\phi_{r_0})$, given that both the fluid viscosity $\eta_f$ and $\phi_{r_0}$ are known. This approximation is reasonable since all the values of $\phi_{r_0}$ under consideration remain low.

The CTS model yields a satisfactory fit with only two adjustable parameters, namely $\sigma_y$ and $\dot{\gamma}_{CTS}^p$, displayed as function of $\phi_{r_0}$ in Fig~\ref{figure1}b-c. In the case of gelling dispersions (\emph{i.e.} $\phi_{r_0} > \phi_{r_0}^{gel}$), the yield stress follows a scaling law $\sigma_y \propto \phi_{r_0}^{3.3}$, consistent with findings in other colloidal gel systems~\cite{Piau1999}. $\dot{\gamma}_{CTS}^*$ roughly follows the same power law, $\dot{\gamma}_{CTS}^* \propto \phi_{r_0}^{3.2}$. Using $\sigma_y$ and $\dot{\gamma}_{CTS}^*$ as scaling factors for the $y-$ and $x-$axes, the flow curves can be rescaled onto a single master curve (Fig.~\ref{figure1}e), provided that $\phi_{r_0} > \phi^{gel}$. This master curve implies a consistent microstructural scenario along the entire flow curve for all volume fractions. Furthermore, the observed relationships $\sigma_y \sim \dot{\gamma}_{CTS}^* \sim \phi_{r_0}^3$ indicates that the initial volume fraction of CB particles completely accounts for both the horizontal and vertical shifts of the flow curves as the volume fraction varies. This is noteworthy given the structural complexity expected for attractive dispersions. In the next section, our objective is to characterize the structure of CB dispersions and to provide insights into how $\phi_{r_0}$ influences the flow characteristics of these dispersions.

\subsection{Microstructure of the carbon black dispersions under shear}

\begin{figure*}
    \includegraphics[scale=0.53, clip=true, trim=0mm 0mm 0mm 5mm]{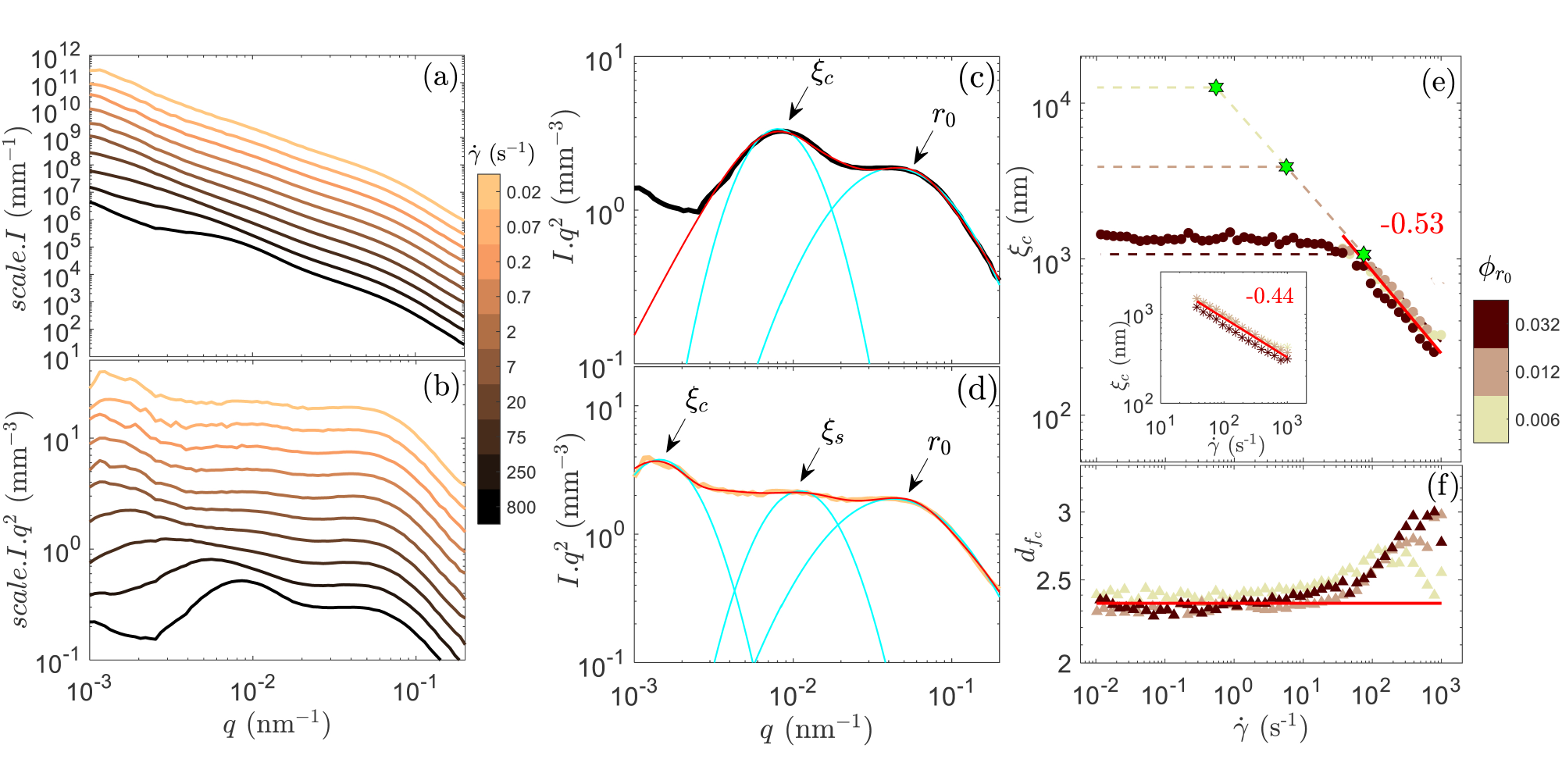}
    \centering
    \caption{Structure of the carbon black dispersion at $\phi_{r_0}$ = 0.032 during a flow curve test. (a) Scattered intensity $I(q)$ $vs$ the wave vector $q$. (b) Kratky representation, $Iq^2$ $vs$ $q$, of the data in (a). The curves are shifted along the $y-$axis for better visualization. (c)-(d) Typical scattering curves in Kratky representation obtained at high ($\dot{\gamma}=800$~s$^{-1}$) and low shear ($\dot{\gamma}=0.02$~s$^{-1}$), respectively. The cyan lines show the fit of the peak positions using Eq.~ \ref{eq:baeza}. The red lines are the modified Beaucage fits using Eq.~\ref{eq:Beaucage1}. (e) Evolution of the cluster size $\xi_c$ with the shear rate $\dot{\gamma}$ for three different $\phi_{r_0}$, determined with the Beaucage model (Eq.~ \ref{eq:Beaucage1}). Note that for $\phi_{r_0}$ = 0.006 and 0.012, the cluster size increases beyond the resolution of the USAXS experiments at low $\dot{\gamma}$. Dash lines are extrapolated cluster sizes and green stars correspond to the cluster size at percolation $\xi_c^*$, determined using the microstructural model (See Fig.~\ref{fig:hydro}). Red line is the best power law fit for $5.10^{2} < \dot{\gamma} < 10^3$~s$^{-1}$. Inset: $\xi_c$ $vs$ $\dot{\gamma}$ determined with the model-free analysis (Eq.~ \ref{eq:baeza}). (f) Fractal dimension of the cluster $d_f$ obtained from the modified Beaucage fit. The red line indicates $d_f = 2.35$, the mean value of $d_f$ for $10^{-2} < \dot{\gamma} < 10^1$~s$^{-1}$.
    } 
    \label{figure3}
\end{figure*}

The structural analysis of CB dispersions was conducted using USAXS at three volume fractions, namely $\phi_{r_0}$ = 0.6, 1.2 and 3.2\%. We recorded the scattering intensity $I(q)$ as function of the scattering wave number $q$ within the range of 1.$10^{-3}$ to 2.$10^{-1}$ nm$^{-1}$ during flow curve measurements. It is noteworthy that $I(q)$ exhibited isotropic behavior for all measurements (see Appendix, Fig.\ref{fig:anisotropy}), ruling out any anisotropic structuring over the probed length scales. This isotropic behavior aligns with previous observations in CB dispersions~\cite{Hipp2021, Dages2022b}. Consequently, our analysis focuses on the azimuthally averaged intensity $I(q)$.

Fig.~\ref{figure3}a presents $I(q)$ at various shear rate values for $\phi_{r_0} = 0.032$. Similar scattering patterns were observed for dispersions at lower volume fractions, and they displayed the same shear rate dependence (see Appendix, Fig.~\ref{fig:beaucage}).
Upon transitioning from high to low shear rates, the scattering curves exhibit two and then three distinct bumps. These features become more pronounced in the Kratky representation ($I(q)q^2$ $vs$ $q$), where the bumps are transformed into peaks (Fig.~\ref{figure3}b). At higher $q$ values (\emph{i.e.} $q \approx 5.10^{-2}$~nm$^{-1}$), the first bump corresponds to the form factor of the CB particles of size $r_0$, as evidenced in Fig.\ref{figure0}. Notably, the position of this bump is independent of the shear rate as shear does not alter the size of the CB particles.

As previously stated, the scattering curves exhibit two characteristic lengths at high shear rates, denoted as $r_0$ and $\xi_c$ (Fig.~\ref{figure3}c-d), and an intermediate length scale $\xi_s$ becomes apparent at lower shear rates (approximately $\dot{\gamma} <$ 75 s$^{-1}$). For a precise and empirical determination $\xi_c$ and $\xi_s$ and their relationship with $\dot{\gamma}$, we conducted fitting procedures on the peaks observed in the Kratky representation (Fig.~\ref{figure3}c-d). These fittings were carried out using a log-normal function~\cite{Baeza2013}:

\begin{equation}
    G_i(q) = \frac{A_i}{\sqrt{2 \pi} \sigma_i q} \exp{\left(-\frac{\ln{(\frac{q}{q_i})^2}}{2\sigma_i^2}\right)} \\  
    \label{eq:baeza}
\end{equation}
where $G_i(q)$ describes a log-normal function of average position $q_i$, width $\sigma_i$ and amplitude $A_i$. The estimated radius ($\xi_i = \pi \slash q_i$) are displayed for $\phi_{r_0}$ = 0.006, 0.012 and 0.032 in Fig.~\ref{fig:Xi_s}b in Appendix. 

\jb{The shear-dependent size $\xi_c$ is attributed to the form factor of clusters, formed through the aggregation of CB particles, consistent with previous reports on CB dispersions~\cite{Dages2022b,Richards2017, Hipp2021}. At high shear rates, typically $\dot{\gamma} = 800$s$^{-1}$, the lack of structure factor at low $q$ indicates that these clusters are independant from each other (Fig.~\ref{figure3}a-b and Fig.~\ref{fig:beaucage} in Appendix).}
Let us first focus on $\phi_{r_0}$ = 0.032. At high shear rates, $\xi_c$ is inversely proportional to the shear rate according to $\xi_c = B \dot{\gamma}^{-m}$, where $B\simeq 10^4$ nm/s$^m$ and $m\simeq0.44$ (inset of Fig.~\ref{figure3}e). Below a critical shear rate of about 50~s$^{-1}$, $\xi_c$ becomes constant with a critical value close to 2~\textmu m. This behavior coincides well with the simulation of Varga et al.~\cite{Varga2018} where a similar scaling with $\dot{\gamma}$ and $m=1/2$ was found. 
The evolution of $\xi_c$ at lower volume fractions ($\phi_{r_0}$ = 0.006 and 0.012) is similar to the one observed at $\phi_{r_0}$ = 0.032. At high shear rates, the data even superimpose. However, at lower shear rates, data are lacking because the peak in $I(q)$ corresponding to $\xi_c$ shifts to values lower than the minimum $q$ accessible in the USAXS experiments. Based on simulations studies~\cite{Varga2018}, for $\phi_{r_0}$ = 0.006 and 0.012, one can expect that the scaling $\xi_c = A \dot{\gamma}^{-m}$ holds down to a critical shear rate lower than 50~s$^{-1}$ and that the cluster size converges to values much larger than 2~\textmu m. For $\phi_{r_0}$ = 0.006 and 0.012, extrapolations of $\xi_c$ $vs$ $\dot{\gamma}$ (Fig.~\ref{figure3}e) will be justified during the microstructural analysis developed in the next section. 

So far, the present results show that the cluster size is governed by the background fluid stress, $\sigma_f = \eta_f \dot{\gamma}$ in agreement with \cite{Sonntag1986, Varga2018}. This is in contrast to the proposition in~\cite{bouthier2023,Hipp2021}, where the dispersion stress was proposed as the determining factor (see Appendix, Fig.~\ref{fig:xic_stress} for an evolution of $\xi_c$ $vs$ $\sigma$). 

$\xi_s$ is found to be independent of the shear rate and slightly diminishes as $\phi_{r_0}$ increases (see Appendix, Fig.\ref{fig:Xi_s}). The exact interpretation of $\xi_s$ remains uncertain. In the context of CB gels at rest, $\xi_s$ had previously been considered either as the separation distance between two neighboring clusters, with the idea that clusters interpenetrate due to the fact that $\xi_c > \xi_s$~\cite{Dages2022b}, or as an intermediate cluster size contributing to the construction of clusters with a size of $\xi_c$~\cite{Bouthier2022b}.  {Alternatively, as proposed in the context of colloidal gelation under quiescent conditions, $\xi_s$ could represent the distance over which particles are transported to the growing cluster. In this case, $\xi_s$ is attributed to a correlation hole corresponding to the size of the depleted region surrounding the clusters~\cite{carpineti1995, narayanan2020}. \tg{Given that $\xi_s$ is smaller that $\xi_c$, it could also represent the depletion at the aggregation length scale. }}

\begin{figure*}[ht]
    \includegraphics[scale=0.52, clip=true, trim=0mm 0mm 0mm 0mm]{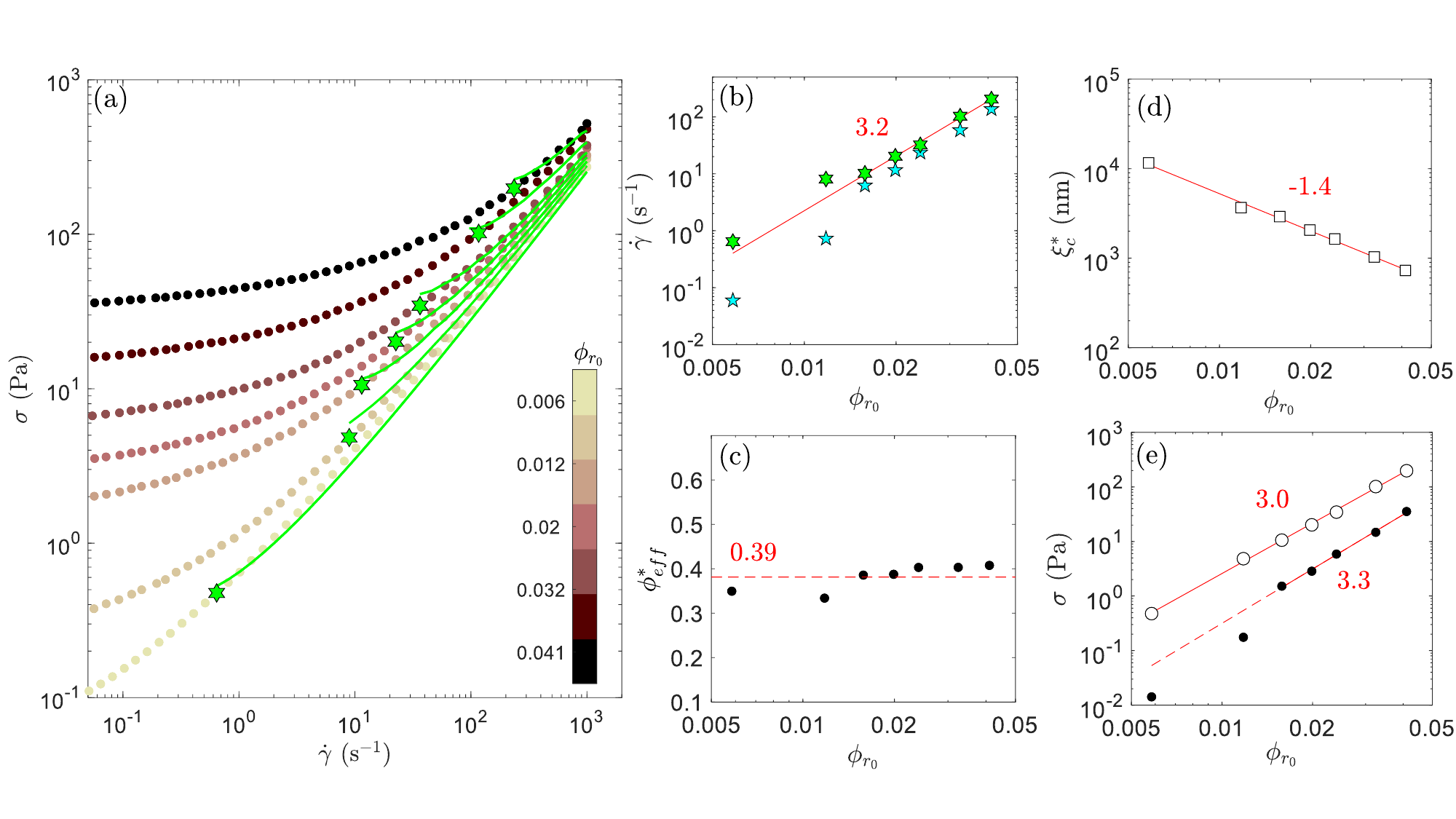}
    \centering
    \caption{Microstructural modeling of the hydrodynamic stress of the flow curve. (a) Comparison between the experimental flow curves (circles) and $\sigma_h$ (green line) determined using Eq.~\ref{eq:phieff} and~\ref{eq:etaKD} and the structural parameters of CB dispersions measured by USAXS. Stars are points of coordinate ($\dot{\gamma}^*$, $\sigma^*$) such that $\sigma_h \approx \sigma$, interpreted as the dynamic percolation of clusters. (b) Comparison between $\dot{\gamma}^*$ (green hexagrams) and $\dot{\gamma}_{CTS}^*$ (cyan pentagrams) as a function of the volume fraction of CB particles $\phi_{r_0}$. (c) Effective volume fraction of clusters at percolation $\phi^*_{\mathrm{eff}}$ displayed as function of $\phi_{r_0}$. Dash line indicates the mean value of $\phi^*_{\mathrm{eff}} \approx 0.39$. (d) Cluster length $\xi_c^*$ at percolation displayed as a function of $\phi_{r_0}$. Red line indicates the best power fit. (e) Comparison between the stress at percolation $\sigma^*$ (empty circles) and the apparent yields stress $\sigma_y$ (solid circles) as a function of $\phi_{r_0}$. Red line indicates the best power law fit.
    }
    \label{fig:hydro}
\end{figure*}

\begin{figure*}
    \includegraphics[scale=0.8, clip=true, trim=0mm 0mm 0mm 0mm]{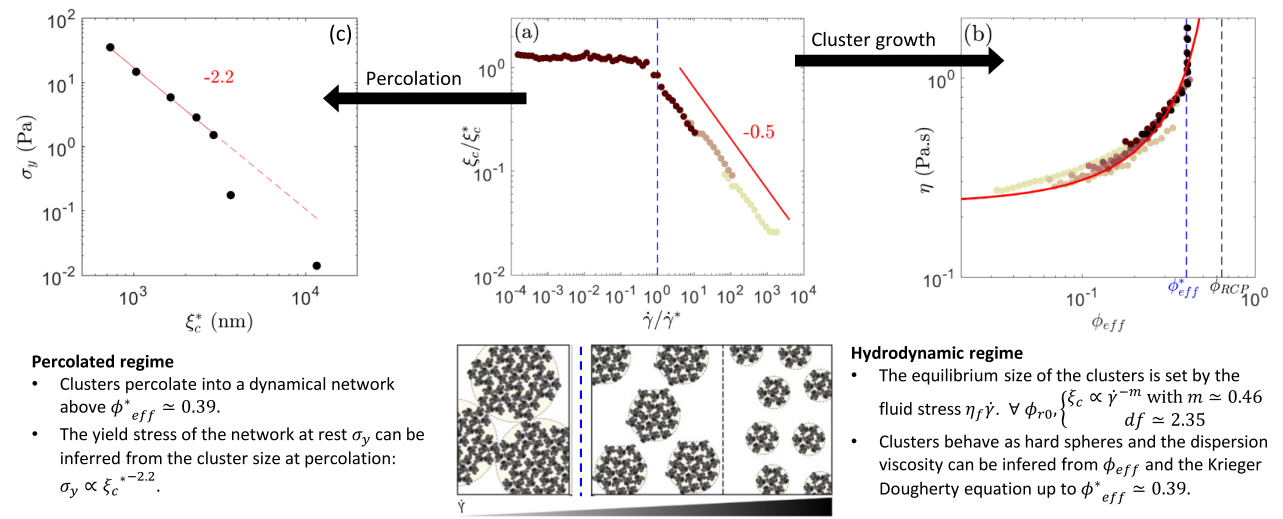}
    \centering
    \caption{Transition between the hydrodynamic regime and dynamic percolation. (a) Normalized cluster size $\xi_c \slash \xi^*_c$ as a function of the normalized shear rate $\dot{\gamma} \slash \dot{\gamma}^*$. $\xi^*_c$ and $\dot{\gamma}^*$ are the cluster size and shear rate at percolation, respectively. (b) Viscosity $\eta$ of the CB dispersions as a funtion of the effective volume fraction of clusters $\phi_{\mathrm{eff}}$ using Eq.~\ref{eq:phieff}. The red line is the viscosity issued from the Krieger-Dougherty model with $\eta_f = 0.234$ Pa.s and $\phi^M_{\mathrm{eff}} = 0.64$. Blue and black dotted lines indicate the volume fraction of clusters at percolation $\phi_{\mathrm{eff}}^*$ and at $\phi^M_{\mathrm{eff}}$, respectively. (c) Yield stress $\sigma_y$ as a function of the cluster size at percolation $\xi^*$ for dispersions with different volume fractions of carbon black particles. The red line represents the best-fitting power law. Bottom: Schematic of the evolution of the CB dispersions microstructure with the shear rate $\dot{\gamma}$.
    }
    \label{fig:KD}
\end{figure*}


To extract the fractal dimension of clusters, the scattering data were fitted with a modified two-level Beaucage model~\cite{Keshavarz2021, Dages2022b}, accounting for the scattering of the particles $r_0$ and of the clusters $\xi_c$ (see Appendix, section E). To account for the intermediate level $\xi_s$, the cluster-level term was multiplied by an empirical structure factor to account for some correlated region~\cite{Keshavarz2021}. The modified Beaucage model provides a good fit of the experimental data for all volume fractions (Fig.~\ref{figure3}c-d and Fig.~\ref{fig:beaucage} in Appendix). Fig.~\ref{figure3}e-f display the cluster sizes and fractal dimensions as function of the shear rate. The cluster sizes obtained from the modified Beaucage model correspond well with the values obtained from the model free analysis, also displayed in Fig.~\ref{figure3}e. The power law fit of $\xi_c$ $vs$ $\dot{\gamma}$ yields $m \simeq 0.53$, confirming that the value of the breaking exponent is about $0.5$. In Fig.~\ref{figure3}f, the fractal dimension shows a non-monotonic evolution with the shear rate. Overall, $d_f$ is close to 3 at high shear rates and tends toward a constant value $d_f$ = 2.35 when $\dot{\gamma}$ decreases. 
Determining $d_f$ at high shear rates presents notable challenges for two primary reasons. First, in $I(q)$ vs. $q$, $d_f$ is determined by the power-law relationship $I(q)\sim q^{-d_f}$ observed between the bumps associated with $r_0$ and $\xi_c$. At high shear rates, where $r_0$ and $\xi_c$ are in close proximity, this power-law behavior extends over a relatively limited range, resulting in insufficient statistical data for accurate determination of $d_f$. Secondly, when examining the structure factor (see Appendix, Fig.~\ref{fig:Sq}), an anticorrelation peak is observed at $q\approx 0.03$~nm$^{-1}$ resulting from attractive interactions among the CB particles. This anticorrelation peak is not accounted by the modified Beaucage model, which leads to an overestimation of $d_f$. At low shear rate, the extended distance between the bumps associated to $r_0$ and to $\xi_c$ allows for a more reliable determination of $d_f$. Moving forward, we disregards the values of $d_f$ obtained at high shear. The fractal dimension is assumed to remain constant and equal to $d_f=2.35$ for all $\phi_{r_0}$ and across the entire range of shear rates ($10^{-2} < \dot{\gamma} < 10^3$~s$^{-1}$), a value which aligns with values predicted by theoretical models~\cite{Zaccone2009,Conchuir2014} and those obtained through other experimental determinations~\cite{Sonntag1986, Hipp2021}.

\tg{Because of the complexity of the SAXS data, the structural analysis proposed here contains many simplifications. Nonetheless, from this approach, we have extracted the general trend of the structuring of CB dispersions under flow: the cluster size is a function of the shear rate so that $\xi_c \propto \dot{\gamma}^{-m}$ with $m$ approximated to $0.5$, and the fractal dimension of clusters is considered to be constant with $d_f = 2.35$. The analysis developed thereafter relies on this approximations and suffice to model the flow curves.}


\subsection{Modeling the hydrodynamic regime of the flow curve based on the cluster structure.}


In the previous section, rheo-USAXS measurements have shown that CB dispersions undergo an hydrodynamic regime at high shear rates. In this regime, for all $\phi_{r_0}$, the cluster size scales as $\xi_c \propto \dot{\gamma}^{-m}$ with $m\approx0.5$ and their fractal dimension remains constant at $d_f \approx 2.35$. For the lowest volume fractions, due to experimental limitations, the cluster size cannot be measured beyond 2~\textmu m and the extent of the hydrodynamic regime is not directly accessible. To address this limitation, the cluster size is extrapolated based on the relationship $\xi_c \propto \dot{\gamma}^{-m}$ observed in Fig.~\ref{figure3}e.  {Using the measured and extrapolated microstructural parameters} of the CB dispersions, an hydrodynamic stress  {$\sigma_h=\eta(\phi_{\mathrm{eff}})/\dot{\gamma}$ is calculated using Eq.~\ref{eq:phieff} for $\phi_{\mathrm{eff}}$ and Eq.~\ref{eq:etaKD} for $\eta$. The scaling factor $k$ in Eq.~\ref{eq:phieff} was calculated~\cite{Ehrl2009} as $k \approx 4.46d_f^{-2.08}$ leading to $k \approx 0.75$. }

As reported in Fig.~\ref{fig:hydro}a, $\sigma_h$ captures well the flow curves down to a critical shear rate $\dot{\gamma^*}$, identified when $\sigma_h$ becomes lager than the dispersion stress $\sigma$. Indeed, extrapolation of the cluster size $\xi_c$ leads to an increase of $\sigma_h$ up to $\sigma_h \ge \sigma$. This point is interpreted as a percolation point where the cluster growth becomes restricted by percolation into a dynamic network, with an effective volume fraction $\phi_{\mathrm{eff}}^*$ and and effective yield stress $\sigma^*$. Consequently, for $\dot{\gamma} \le \dot{\gamma^*}$, the cluster size is assumed to become constant and equals at $\xi_c^*$, as depicted in Fig.~\ref{figure3}e where $\dot{\gamma}^*$ is pointed out by stars. This analysis is consistent with the scattering data at $\phi_{r_0}=0.032$, where the cluster sizes at low shear rates are available. 
Fig.~\ref{fig:KD}b displays the dispersion viscosity $\eta$ vs $\phi_{\mathrm{eff}}$, the effective volume fraction of clusters. The collapse of viscosity onto a master curve confirms that the rheology of the CB dispersions depends on the shear dependent volume fraction of clusters. The Krieger-Dougherty fit (red line) accurately describes the dispersions behavior up to approximately $\phi_{\mathrm{eff}}^*\simeq0.4$, which sets the upper limit in effective volume fraction of the hydrodynamic regime.



\subsection{Properties of the percolation point ($\dot{\gamma}^*$, $\sigma^*$).}


The values of $\dot{\gamma}^*$, $\xi_c^*$, $\sigma^*$ and $\phi_{\mathrm{eff}}^*$ are plotted as function of $\phi_{r_0}$ in Fig.~\ref{fig:hydro}b-e. As mentioned in the previous section, these parameters relate of the structure of dispersions when clusters are assumed to percolate into a dynamic network.

In Fig.~\ref{fig:hydro}b, $\dot{\gamma}^*$, obtained from the microstructural approach and $\dot{\gamma}_{CTS}^*$, obtained from the CTS model, are overlaid for $\phi_{r_0}>\phi_{r0}^{gel}$. Thus, $\dot{\gamma}_{CTS}^*$, used to rescale the flow curves in Fig.~\ref{figure1}, is now identified as the shear rate at witch clusters percolate. In Fig.~\ref{fig:hydro}c, $\phi_{\mathrm{eff}}^* \approx 0.39$ is constant for all $\phi_{r_0}$. This value is lower than the random close packing fraction of hard spheres, as illustrated on Fig.~\ref{fig:KD}. Percolation below random close packing has been previously reported for non-Brownian dispersions with adhesive constrains at an "adhesive close packing" fraction $\phi_{acp}=0.55$~\cite{Richards2021}. Varga et al.~\cite{Varga2018} have proposed that a critical $\mathrm{Mn}$ value for a percolated network could be obtained as $\mathrm{Mn}_c\propto \phi_{r_0}^{2 \slash (3-d_f)}$. Assuming that $\mathrm{Mn} \propto \dot{\gamma}$, the critical shear rate $\dot{\gamma}^*$ should scales with $\phi_{r_0}$ with a power exponent $2 \slash (3-d_f) \simeq 3.1$ given that $d_f=2.35$. This is consistent with the observed scaling $\dot{\gamma}^* \propto \phi_{r_0}^{3.2}$ observed in Fig.~\ref{fig:hydro}b.

As shown in Fig.~\ref{fig:hydro}d, the cluster size at percolation $\xi_c^*$ decreases with $\phi_{r_0}$ as $\phi_{\mathrm{eff}}^*$ is reached at higher shear for the highest volume fractions. In fractal aggregation theories~\cite{Shih1990,Wu2001}, it is assumed that the gel network is built of closely packed clusters at $\phi_{\mathrm{eff}}^* = 1$ which directly yields $\xi_c \propto \phi_{r_0}^{1 \slash (d_f-3)}$. For CB dispersions, $\phi_{\mathrm{eff}}^* \ne 1$ but $\phi_{\mathrm{eff}}^*$ is constant and the observed scaling $\xi_c^*  \propto \phi_{r_0}^{-1.4}$ aligns well with $1 \slash (d_f-3) \approx -1.5$ given that $d_f = 2.35$ (Fig.~\ref{fig:hydro}d). Fig.~\ref{fig:hydro}e compares the dependency of $\sigma_y$, measured in the zero shear rate limit, and $\sigma^*$ on $\phi_{r_0}$. Interestingly, both quantities exhibit a similar scaling pattern with respect to $\phi_{r_0}$. This observation supports the idea that the value of $\sigma_y$ is a direct outcome of the percolation point. Indeed, when $\sigma_y$ is plotted against $\xi_c^*$ (Fig.~\ref{fig:KD}(c)), it is found that $\sigma_y \propto {\xi_c^*}^{-2.2}$ a close match to what one would expect~\cite{Bouthier2022a}: $\sigma_y \propto {\xi_c}^{-2}$ for a network of characteristic size $\xi_c$. It is worth mentioning that, when the measuring time $\Delta t$ of the flow curve increases (see Fig.~\ref{fig:FCtime} in Appendix), this direct dependence between $\sigma_y$ and $\xi_c^*$ is probably lost as structural rearrangements may occur at low shear rates.


To sum up, in Fig.~\ref{figure1} the flow curve $\sigma$ $vs$ $\dot{\gamma}$ of CB dispersions at various $\phi_{r_0}$ have been rescaled using a critical shear rate $\dot{\gamma}^*_{CTS}$ and $\sigma_y$ as scaling factors for the $x-$ and $y-$axis, respectively. Now, the origin of this scaling can be understood on a microstructural basis. The critical shear rate $\dot{\gamma}^*_{CTS}$ corresponds to the dynamic percolation of clusters at $\phi_{\mathrm{eff}}^* \simeq 0.39$. The decrease of the cluster size $\xi_c^*$ at percolation is then responsible for the increase of $\sigma_y$ with $\phi_{r_0}$, as the network structure in the zero shear limit is inherited from the percolation point. When $\phi_{r_0}$ approaches $\phi^{gel}_{r_0}$, dispersions no longer displays a yield stress and the structuring at low shear is somewhat different. The percolation point, defined by $\dot{\gamma}^*$, $\xi_c^*$, $\sigma^*$, and $\phi_{\mathrm{eff}}^*$, holds significant importance in understanding the flow behavior of CB particles, being the transition point between the hydrodynamic regime and the dynamic percolation of clusters. In Fig.~\ref{fig:KD}a, $\xi_c$ $vs$ $\dot{\gamma}$ at various $\phi_{r_0}$ can be normalized onto a master curve using the percolation point coordinates, \emph{i.e.} $\dot{\gamma}^*$ and $\xi_c^*$, as scaling factors for the $x-$ and $y-$axis, respectively. The experimental results display on Fig.~\ref{fig:KD}a are fully consistent with simulations performed in~\cite{Varga2018}.


\section{Conclusion and perspectives}
\label{s:discuss}
A series of rheo-USAXS experiments was conducted on CB dispersions with $\phi_{r_0} \in [0.006, 0.041]$ to investigate the response of attractive colloidal dispersions to mechanical shear. These experiments involved rapidly changing the shear rate from high to low values, revealing a microstructural scenario depicted in Fig.~\ref{fig:KD}. 

In the hydrodynamic regime, attractive CB particles assemble into clusters with a fractal dimension of $d_f=2.35$, and their size inversely scales with the shear rate as $\xi_c\propto\dot{\gamma}^{-m}$, where $m$ is approximately equal to 0.5. The size of these clusters is primarily determined by the fluid stress, represented by $\sigma=\eta_f \dot{\gamma}$, where $\eta_f$ is the fluid background viscosity. The viscous contribution of the clusters can be approximated by the one of a hard sphere dispersion and the viscosity of CB dispersions can be calculated using the Krieger-Dougherty equation, provided we replace the CB particle volume fraction $\phi_{r_0}$ by an effective volume fraction of clusters $\phi_{\mathrm{eff}}$. This hydrodynamic regime persists down to a critical point characterized by coordinates ($\dot{\gamma}^*$, $\sigma^*$). \jb{Although this model capture well the interplay between structure and rheology, the dynamics underlying this process remain to be fully understood. Specifically, what mechanisms drive cluster growth as the shear rate decreases? It might involves a balance between the aggregation of clusters upon collision and a feedback erosion process driven by the shear rate that restricts cluster size~\cite{Zaccone2011}.}

At this juncture point ($\dot{\gamma}^*$, $\sigma^*$), the cluster growth stops and their effective volume fraction stabilizes at $\phi_{\mathrm{eff}}\approx 0.39$, irrespective of $\phi_{r_0}$. We propose that at $\phi_{\mathrm{eff}}^*$, clusters percolate into a dynamic network. This hypothesis gains support from the observation that $\sigma^*$ closely tracks variations in the dynamical yield stress $\sigma_y$. When $\phi_{r_0}$ increases, the clusters size at percolation decreases, leading to a higher $\sigma_y$. In essence, the properties of dispersions in the limit where $ \dot{\gamma} =0$ are inherited from the percolation point. As a result, the dynamic percolation point is a key parameter to explain the mechanical response of CB dispersions under shear. While the cluster growth in the hydrodynamic regime proceeds independently of the volume fraction, the percolation of clusters at a constant $\phi_{\mathrm{eff}}^*$ determines both the critical shear rate and the yield stress of the flow curve.

Microstructural models have traditionally partitioned the stress response of attractive dispersions into an hydrodynamic and an elastic contributions~\cite{DeRooij1994, Potanin1995} and a robust and self-consistent comprehension of these two regimes has been established for CB dispersions. However, in Fig.~\ref{fig:hydro}a, it is clear that the hydrodynamic regime does not extent to low shear rates, in the vicinity of $\sigma_y$, and that an additional contribution is required to model the flow curve at intermediate shear rates ($\dot{\gamma}<\dot{\gamma}^*$). Previously, we have stressed that the cluster size at percolation was maintained when $\dot{\gamma}$ is further decreased. It pictures the idea that, during the intermediate regime, the structure of the dispersion is shear-independent and composed of large and crowded clusters, comparable with the flow of adhesive non-Brownian dispersions. Based on the phenomenological model of Wyart and Cates~\cite{Wyart2014}, constraint-based rheology was shown to successfully model all types of flow curves~\cite{Guy2018}. In this framework, particle sliding and rolling are constrained by adhesive and/or frictional contacts that are accounted by a shear-dependent jamming point $\phi_m$. For adhesive non-Brownian dispersions, experimental results ~\cite{Richards2020} highlight the coupling between adhesion and friction to account for the shear-thinning and yield stress of the dispersions. In the case of attractive Brownian dispersions, for $\dot{\gamma} < \dot{\gamma}^*$, the flow of sticky clusters may be successfully modeled using constraint-based rheology. The challenge probably lies in rationalizing the contribution of adhesion and friction in these systems. An alternative method for modeling this plastic regime would involve describing the elastic stress contribution of the dynamic network with a network percolation loss rate that decreases with respect to a characteristic time of $1/\dot{\gamma}$.

Additional research is needed to characterize the intermediate regime. Colloidal systems that are optically transparent would be particularly advantageous, as they would allow for the characterization of mesoscopic structures (with $\xi >$ 2 \textmu m) using optical microscopy or dynamic light scattering. \jb{Besides, measuring the viscoelastic properties of the suspensions during shear using orthogonal superposition rheometry~\cite{Wang2022} may bring further insights to the shear stress contribution during flow curve tests.}
Finally, it would be beneficial to validate the approach used in this paper and compare our findings with those from other attractive colloidal systems subjected to flow. Given the thixotropic behavior of Carbon black dispersions and attractive colloidal dispersions in general, it would also be interesting to study the impact of the flow curve protocol in such frameworks. 
\jb{We have indeed opted to continuously decrease the shear rate. If we were to continuously decrease the shear stress instead, would the control parameter still be the shear rate? Alternatively, one could perform discrete jumps from a higher shear rate to a lower one. In this scenario, would the cluster size be determined by the initial or final value of the shear rate? It is observed that at lower shear rates, the viscosity decreases over time~\cite{Wang2022}. Is this behavior attributed to slip at the boundaries of the Couette cell or to the restructuring of the clusters and the network over time?}

\section*{Author Contributions}
JB and TG carried out the experiments. JB analysed the data. TG, LVB and AP helped JB analyzing the data. TG and JB wrote the paper. TG designed and managed the project.
 
\section*{Conflicts of interest}
There are no conflicts to declare.
\section*{Acknowledgements}
The authors are especially grateful to the ESRF for beamtime at the beamline ID02 (proposal SC-5236) and Theyencheri Narayanan for the discussions and technical support during the USAXS measurements. The authors thank Szilvia Karpati (ENS de Lyon) and the CIQLE (Centre d'Imagerie Quantitative Lyon-Est) microscopy platform for their help with the electron microscopy imaging of the CB particles. 
The authors express their appreciation for valuable discussions held with Thibaut Divoux and Sébastien Manneville.
This work was supported by the Région Auvergne-Rhône-Alpes ``Pack Ambition Recherche", the LABEX iMUST (ANR-10-LABX-0064) of Université de Lyon, within the program "Investissements d'Avenir" (ANR-11-IDEX-0007), the ANR grants (ANR-18-CE06-0013 and ANR-21-CE06-0020-01). This work benefited from meetings within the French working group GDR CNRS 2019 ``Solliciter LA Matière Molle" (SLAMM).


\section{Appendix}


\subsection{Effect of the duration on the flow curve}
\begin{figure}[b]
    \includegraphics[scale=0.45, clip=true, trim=8mm 8mm 0mm 0mm]{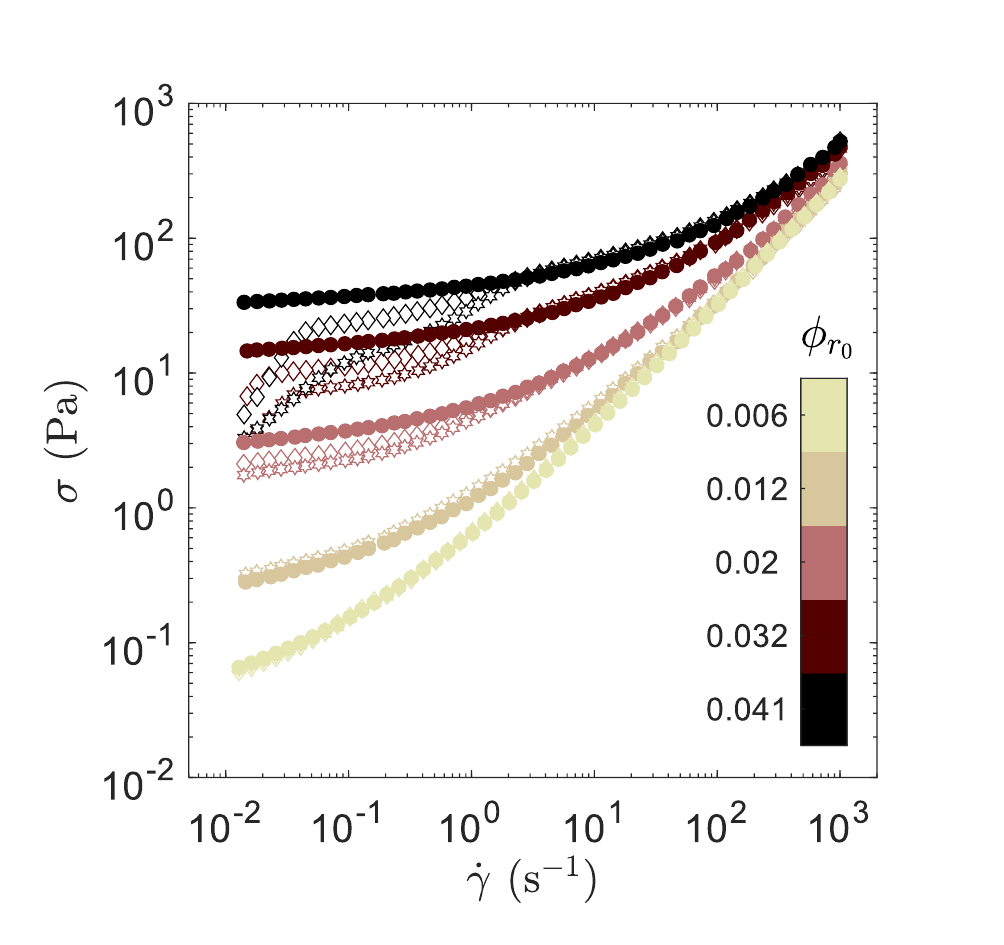}
    \centering
    \caption{Flow curve of carbon black dispersions at various volume fraction $\phi_{r_0}$. The ramp was conducted with $\Delta t = 1$ (circle), $50$ (diamond) and $100$ (star) s/pts.}
    \label{fig:FCtime}
\end{figure}

As shown in Fig.~\ref{fig:FCtime}, the shape of the flow curve depends on the shear duration $\Delta t$ applied to measure each point. Two regimes can be distinguished depending on the shear rate: at high shear rate ($\dot{\gamma} > 1$~s$^{-1}$), the stress response is independent on $\Delta t$ while at low shear rate ($\dot{\gamma} < 1$~s$^{-1}$) and for high volume fractions of CB particles $\phi_{r_0}$, a drop of the stress can be observed. The drop is all the more important that $\Delta t$ is large. 

\subsection{Herschel-Bulkley model}
\begin{figure*}
    \includegraphics[scale=0.47, clip=true, trim=0mm 0mm 0mm 0mm]{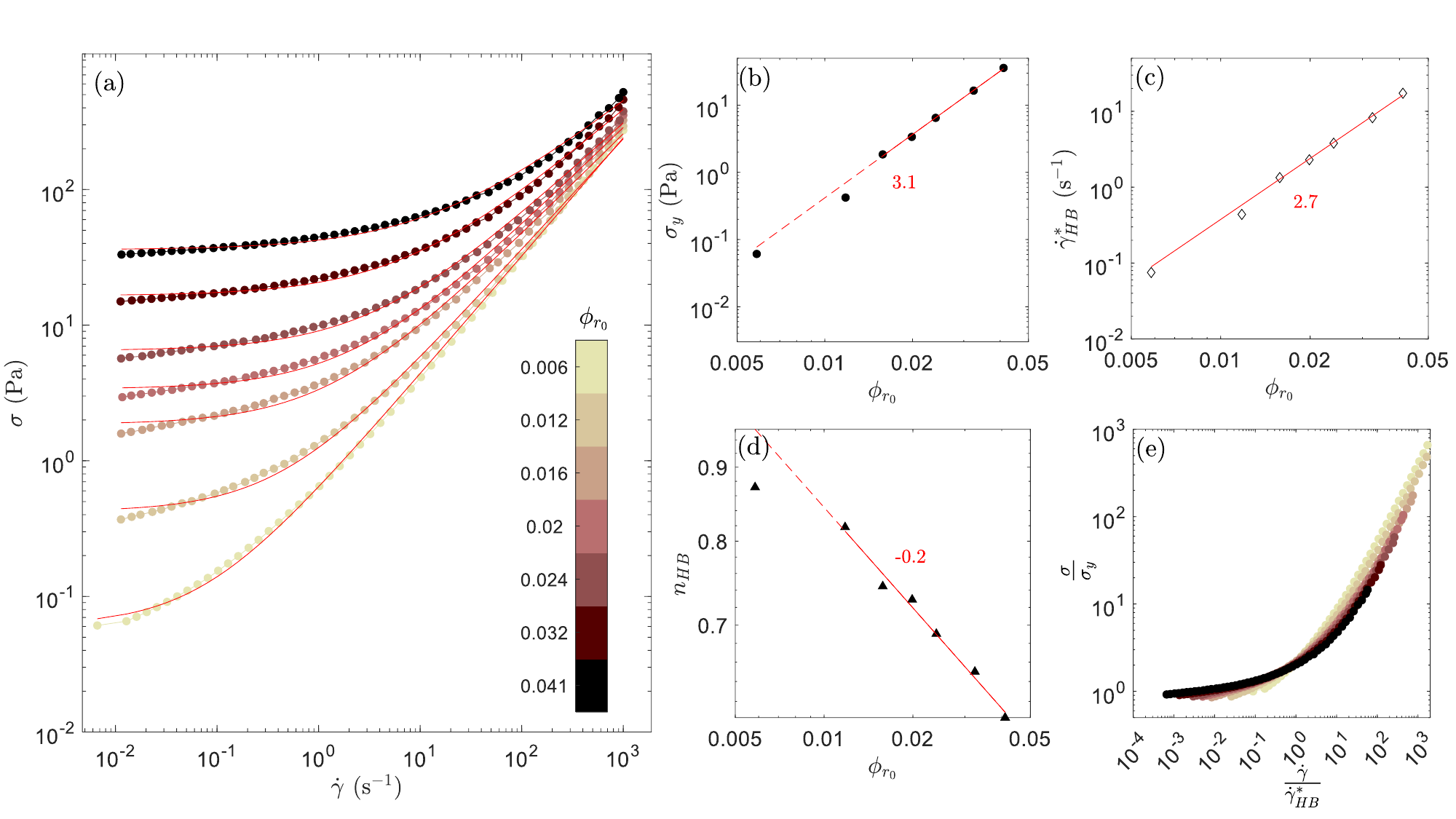}
    \centering
    \caption{Herschel-Bulkley model of the carbon black flow curves. (a) Shear stress $\sigma$ $vs$ shear rate $\dot{\gamma}$ during fast flow curves of carbon clack dispersions at various volume fraction of CB particles $\phi_{r_0}$. Inset displays the dynamic viscosity $\eta$ $vs$ $\dot{\gamma}$. The red curves correspond to the best fits with the Herschel-Bulkley component model (eq.~\ref{eq:hb}). (b)-(d) Evolution of the parameters of the Herschel-Bulkley model with $\phi_{r_0}$. $\sigma_y$ is the apparent yield stress, $\dot{\gamma}^*_{HB}$ is a critical shear rate and $n$ is the power exponent. Red lines are the best power law fit of the data. (e) Normalized flow curves: $\sigma \slash \sigma_y$ $vs$ $\dot{\gamma} \slash \dot{\gamma}^*_{HB}$.}
    \label{fig:HB}
\end{figure*}

In Fig.~\ref{fig:HB}, the Herschel-Bulkley model provides an acceptable fit of the data. However the Herschel-Bulkley model is less efficient that the CTS model to capture the data at intermediate shear rate, around $\dot{\gamma}\approx 1$~s$^{-1}$. Notably, the critical shear rate $\dot{\gamma}^*_{HB}$ shows a different scaling with $\phi_{r_0}$ in comparison with $\dot{\gamma}^*$ obtained from the CTS model (Fig.~\ref{fig:HB}c). As pointed in~\cite{Caggioni2020}, the CTS model gets rid of the nonphysical power exponent $n$ (Fig.~\ref{fig:HB}d) by introducing a high shear viscosity $\eta_0$ (Eq.~\ref{eq:trappe}).

\subsection{Form factor}

The form factor of the CB particles was measured with SAXS on a diluted sample ($\phi_{r_0} = 10^{-4}$) at rest and fitted using different models. 

\begin{figure*}
    \includegraphics[scale=0.52, clip=true, trim=0mm 0mm 0mm 0mm]{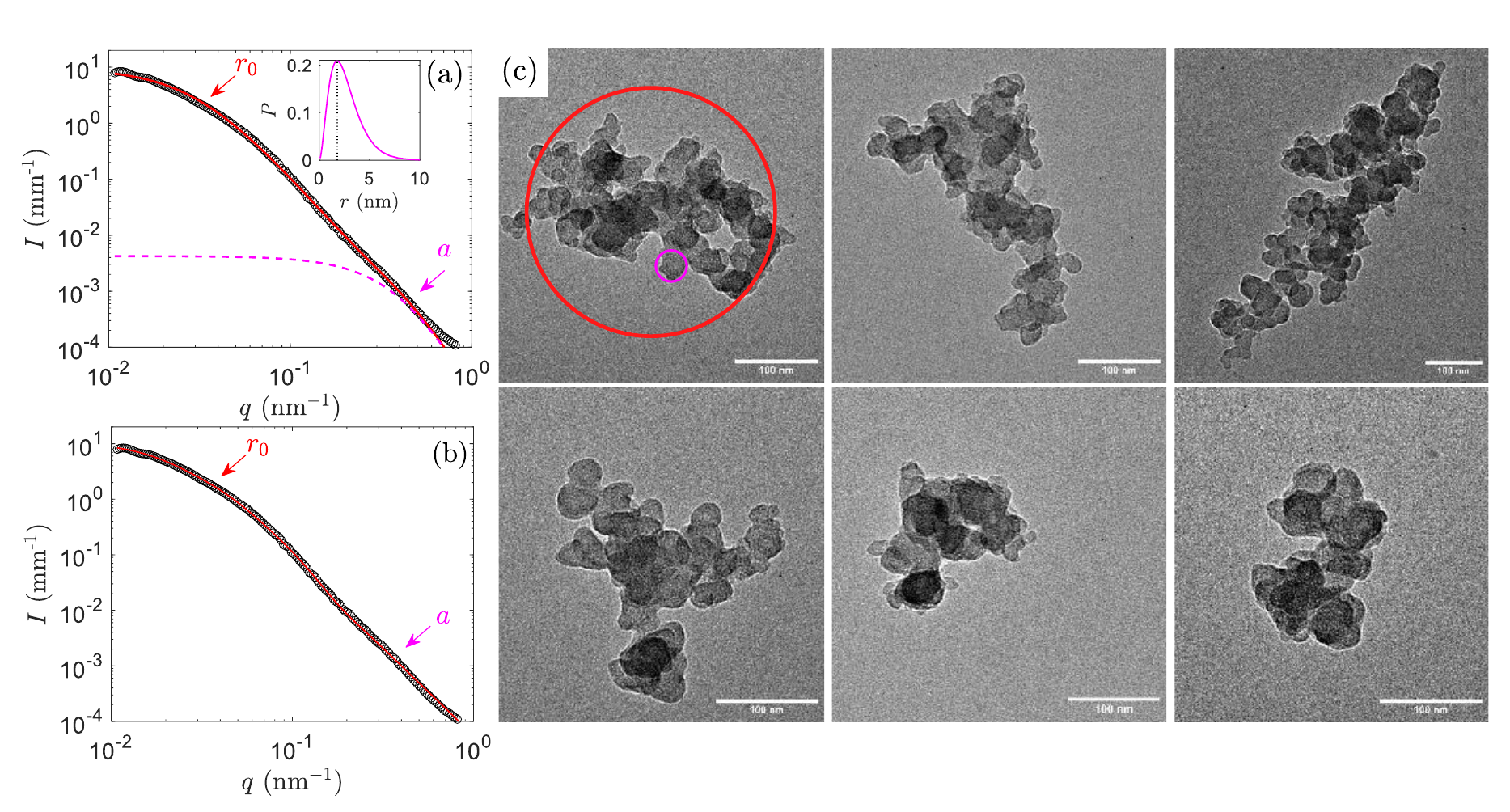}
    \centering
    \caption{Characterization of the Vulcan\textsuperscript{\textregistered}PF carbon black (CB) particles. (a) Fit of the CB form factor using a mass fractal model.  Inset: log-normal distribution of $a$. (b) Fit of the CB form factor using a Beaucage model. 
        Form factor of the CB particles measured by SAXS for $\phi_{r_0}$ = 10$^{-4}$. CB particles are composed of nodules of radius $a$ that are fused to form primary aggregates of radius $r_0$. (c) Representative set of TEM images of individual CB particles.}
    \label{fig:FF}
\end{figure*}

In Fig.~\ref{fig:FF}a, the form factor of the CB particles is described as a combination of a particle form factor $P(q)$, accounting for the nodules, and a fractal structure factor $S(q)$, accounting for the primary aggregates, with $I(q) \propto P(q)S(q)$. $P(q)$ is modeled as the form factor of polydisperse spheres with a log-normal distribution~\cite{Aragon1976} and $S(q)$ is modeled as a mass fractal structure factor~\cite{Teixeira1988}.  The fit yields $a = 3$~nm, $r_0 = 33$~nm, $d_{f_{r_0}}$ = 2.85. The radius of gyration of a fractal aggregate is given by~\cite{Teixeira1988}: $r_0^2 = \frac{d_f(d_f+1)\xi_{r_0}^2}{2}$ which yields $r_0 = 76$~nm. 

In Fig.~\ref{fig:FF}b, the form factor of the CB particles is modeled using a two-level Beaucage model. The first level models the particles of size $a = 20$~nm. The second level models the primary aggregate of size $\xi_{r_0} = 85$~nm and fractal dimension $d_{f_{r_0}}$ = 2.78. 

In both cases, the fit results compares well with the CB particles TEM images in Fig.~\ref{fig:FF}c. 
The characterization of the Vulcan PF particles presented here confirms that these particles are small and compact in comparison with other CB particles previously reported (e.g. $R_g \approx$ 180 nm for Vulcan XC-72 \cite{Richards2017, Dages2021}).

\subsection{Anisotropy}

\begin{figure}
    \includegraphics[scale=0.45, clip=true, trim=0mm 0mm 0mm 0mm]{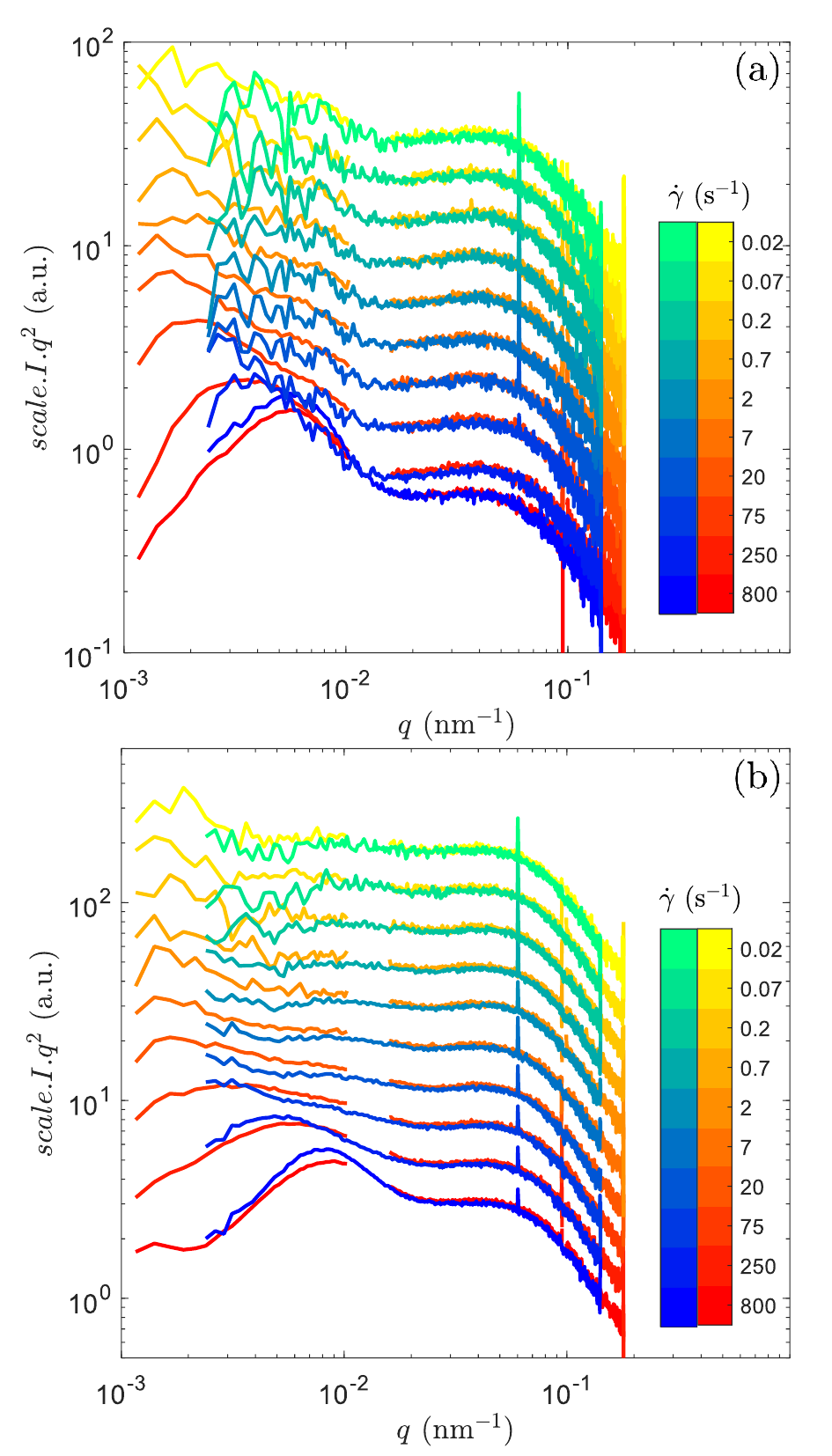}
    \centering
    \caption{Comparison of the scattering intensity between the flow direction (red-yellow) and the vorticity direction (blue-green) at $\phi_{r0}=0.006$ (a) and 0.032 (b). The color gradients code for the shear rate.
    }
    \label{fig:anisotropy}
\end{figure}

Fig.~\ref{fig:anisotropy} displays the scattering curves of CB dispersions, azimuthally averaged either in the flow direction or in the vorticity direction. For the blue-green curves, corresponding to the vorticity direction, the low $q$ region is reduced because of the rectangular shape of the beamstop used. At all shear rates, the superimposition of the curves in both directions indicates the absence of anisotropic structure, also reported by~\cite{Hipp2021} for CB dispersions in oil.

\subsection{Modified Beaucage model}
\label{sb:beaucage}

The intensity spectrum $I(q)$ may be fitted by a modified two-level Beaucage model~\cite{Beaucage1995,Beaucage1996,Keshavarz2021,Dages2022b,bouthier2023} through: 

\begin{equation}
    I\left(q\right)=I_c\left(q\right) S_S\left(q\right) + I_{r_0}\left(q\right)
\end{equation}
with
\begin{equation}
\begin{array}{l}
\left\{
\begin{array}{ll}
I_c(q) & = G_c \mathrm{e}^{\left(-\frac{q^2 \xi_c^2}{3}\right)} + B_c \mathrm{e}^{\left(-\frac{q^2 r_0^2}{3}\right)}\mathrm{erf}\left(\frac{q \xi_c}{\sqrt6}\right)^{-3d_f}   q^{-d_f} \\
S_s(q) & =1+C_s \left( \left(\frac{q\xi_s}{2\pi}\right)^2 + \left(\frac{2\pi}{q\xi_s}\right)^2 \right)^{-1} \\
I_{r_0}(q) & =G_{r_0} \mathrm{e}^{\left(-\frac{q^2 r_0^2}{3}\right)} + B_{r_0}\mathrm{erf}\left(\frac{q r_0}{\sqrt6}\right)^{-3d_{r_0}}q^{-d_{r_0}}
\end{array}
\right.
\end{array}.
\label{eq:Beaucage1}
\end{equation}

\begin{figure*}
    \includegraphics[scale=0.45, clip=true, trim=0mm 0mm 0mm 0mm]{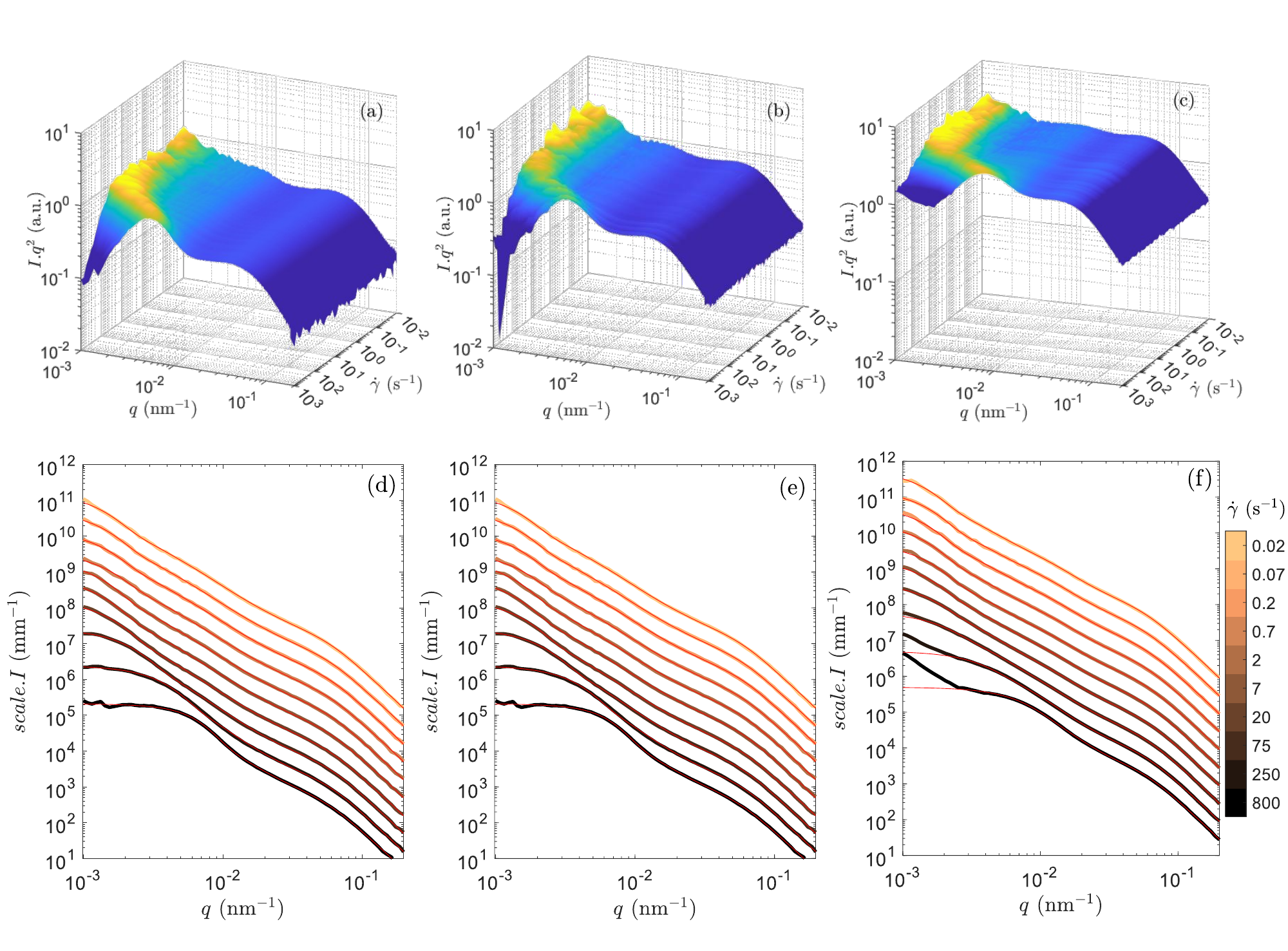}
    \centering
    \caption{(a)-(c) 3D visualization of the shear dependence of the scattering intensity in Kratky representation $Iq^2$ $vs$ $q$ for the volume fractions $\phi_{r0}=0.006$ (a), 0.012 (b) and 0.032 (c). The color gradient codes for the intensity of $Iq^2$. (d)-(f) $I(q)$ and their fit (red lines) using the modified Beaucage model (Eq.~\ref{eq:Beaucage1}) for the volume fractions $\phi_{r0}=0.006$ (d), 0.012 (e) and 0.032 (f). The color gradient from light orange to black codes for the shear rate. $I(q)$ are shifted along the $y-$axis for better visualization.
    }
    \label{fig:beaucage}
\end{figure*}

This model accounts for the presence of three distinct bumps observed in $I(q)$ corresponding to the length scales $r_0$, $\xi_s$, and $\xi_c$ arranged in ascending order. The two-level Beaucage model combines the scattering contribution $I_c(q)$ from the clusters with a size $\xi_c$ and a fractal dimension $d_f$ with the contribution $I_{r_0}(q)$ from the CB particle of size ${r_0}$. To account for the intermediate length scale $\xi_s$, revealed at low shear rate, the cluster intensity $I_c(q)$ is multiplied by an ad-hoc structure factor $S_s(q)$, resulting in an increase in scattering at intermediate $q$. This structure factor $S_s(q)$ is a function that exhibits a peak at $q_s=2\pi/\xi_s$ and reaches a maximum value of $1+C_s/2$. Away from $q_s=2\pi/\xi_s$, it gradually converges to 1. While this choice of $S_s(q)$ is straightforward, it is not entirely satisfactory as it fails to meet thermodynamic limits. Specifically, $S_s(q\rightarrow 0)$ equals 1, whereas it should be proportional to the isothermal compressibility. In Fig.~\ref{fig:beaucage}, we show the fit of the scattering intensity of the CB dispersions using the modified two-level Beaucage model.

\tg{Hipp et al.~\cite{Hipp2021} also study the formation of sheared clusters of carbon black particles. They have characterized the cluster properties using small angle neutron scattering (SANS) and fit their data with the following model:}

\begin{equation}
    I\left(q\right)= P\left(q\right) S_1\left(q\right)S_2\left(q\right)\\
\end{equation}
with
\begin{equation}
\begin{array}{l}
\left\{
\begin{array}{ll}
P(q) & = \left\{
\begin{array}{lll}
\frac{G}{q^s}exp\bigg(\frac{-q^2Rg^2}{3-s}\bigg) \qquad q \leq q_1, \\
\frac{D}{q^m} \qquad \qquad \qquad \ \ q \geq q_1 \\
\end{array}
\right.
\\
S_i(q) & = 1 + \frac{d_f\Gamma(d_f-1)}{[1 + 1/(q\xi)^2]^{(d_f-1)/2}}\frac{sin[(d_f-1)tan^{-1}(q\xi)]}{(qR_0)^{d_f}} \\
\end{array}
\right.
\end{array}.
\end{equation}

\begin{figure}[H]
    \includegraphics[scale=0.45, clip=true, trim=0mm 0mm 0mm 0mm]{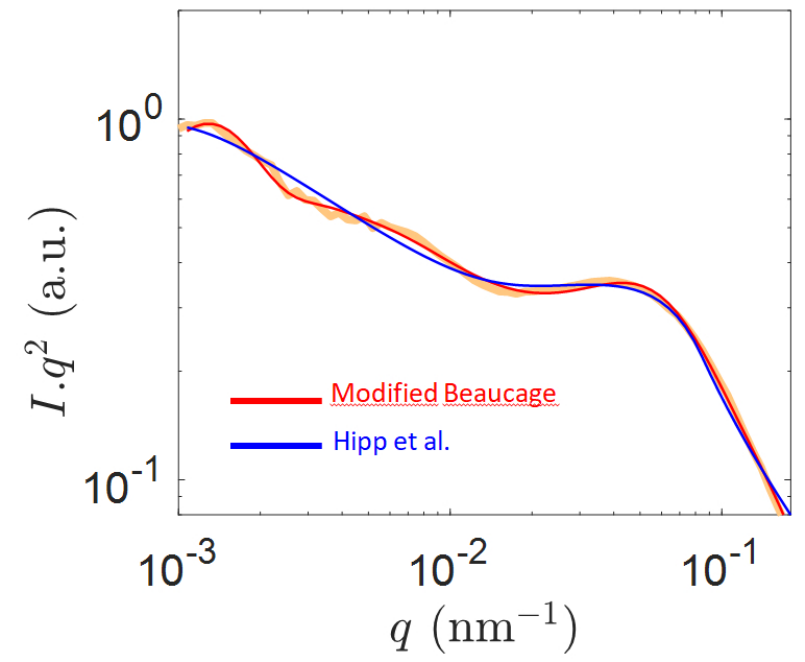}
    \centering
    \caption{Comparison between the modified Beaucage model and the Hipp et al. model.}
    \label{fig:hipp}
\end{figure}

\tg{where $P(q)$ is the form factor of the nodule (Guinier Porod) of size $a$. $S_1(q)$ is a mass fractal model~\cite{Teixeira1988} used to fit the form factor of the CB particles as insecable fractal cluster of size $r_0$ and fractal dimension $d_{fr_0}$ composed of spheres of size $a$. $S_2(q)$ is a mass fractal model~\cite{Teixeira1988} used to fit fractal clusters of size $\xi_c$ and fractal dimension $d_f$ composed of the CB particles.
Although the CB particles and the solvent used by Hipp et al. differ from ours, we attempted to fit our data with their model. In contrast to the modified two-level Beaucage model, the Hipp et al. model does not incorporate the intermediate length scale $\xi_s$. As depicted in Fig.~\ref{fig:hipp}, the Hipp et al. model does not provide a good fit to the SAXS data, it misses by construction $\xi_s$ and tends to underestimate $\xi_c$. This justifies our preference for the modified Beaucage model over the model proposed by Hipp et al.~\cite{Hipp2021}}

\subsection{Evolution of $\xi_c$ with $\sigma$}

\begin{figure}[H]
    \includegraphics[scale=0.45, clip=true, trim=0mm 0mm 0mm 0mm]{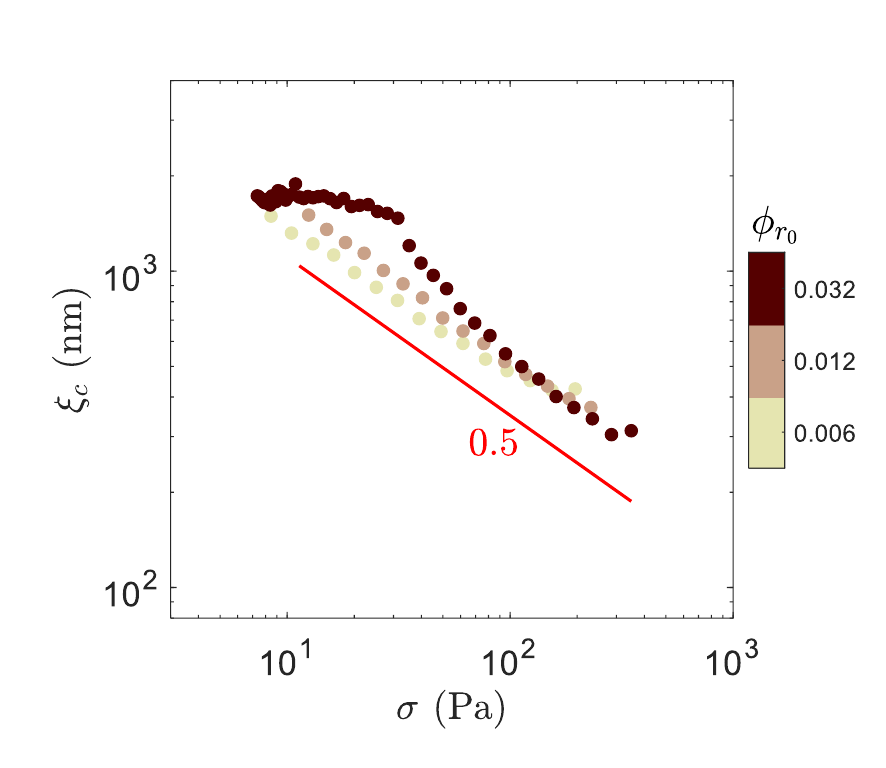}
    \centering
    \caption{Evolution of $\xi_c$ with $\sigma$ for $\phi_{r_0}=0.006$, 0.012 and 0.032. Red line is a power law $\xi \propto \sigma^{-0.5}$ displayed for comparison.}
    \label{fig:xic_stress}
\end{figure}

In Fig.~\ref{fig:xic_stress}, we observe that $\xi_c$ does not scale with $\sigma$.

\subsection{Structure factor}
The structure factor of the CB dispersion $S(q,\phi_{r_0})=\frac{I(q,\phi_{r_0})}{P(q)}\frac{\phi_{r_0}^P}{\phi_{r_0}}$ is obtained by dividing the the scattering intensity of the CB dispersion at $\phi_{r_0}$ by the CB particles form factor measured at $\phi_{r_0}^P=10^{-4}$ as shown in Fig~\ref{fig:FF}. $S(q)$ displays an  anticorrelation peak at $q\sim 0.03$~nm$^{-1}$, associated with attractive interactions among the CB particles the background oil.

\begin{figure}
    \includegraphics[scale=0.45, clip=true, trim=0mm 0mm 0mm 0mm]{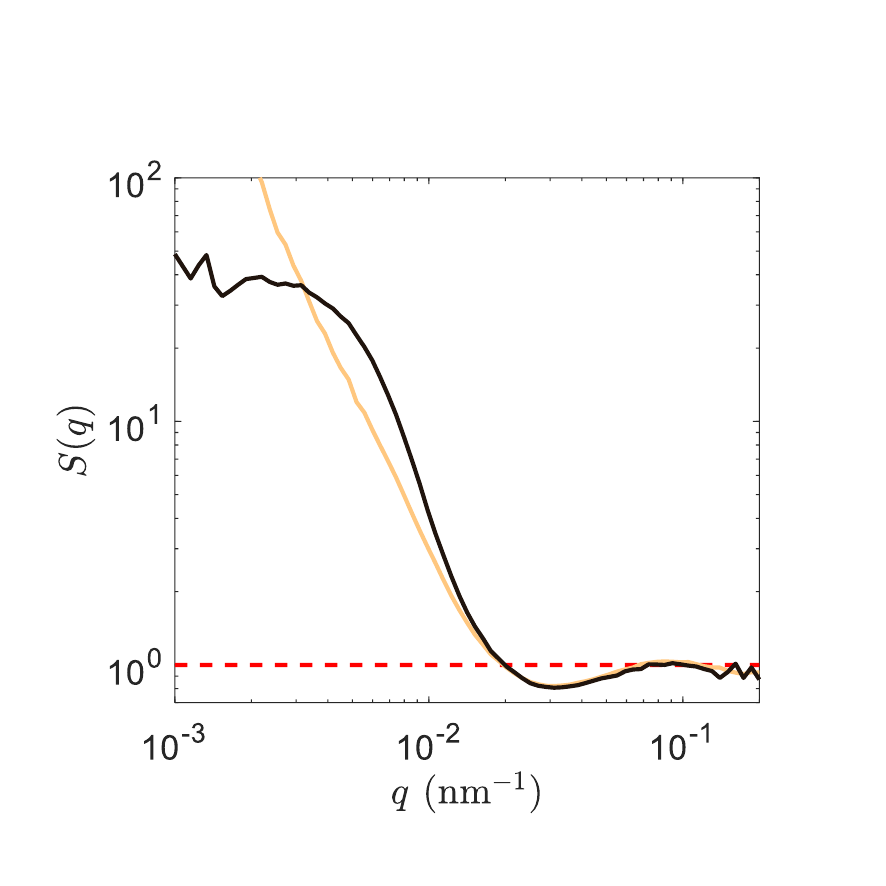}
    \centering
    \caption{Comparison of the structure factor $S(q)$ of the CB dispersion at $\phi_{r_0}=0.006$ at $\dot{\gamma}=800$ (black) and 0.02~s$^{-1}$ (light orange).}
    \label{fig:Sq}
\end{figure}

\subsection{Evolution of $\xi_s$ and $\xi_c$ with $\phi_{r_0}$ and $\dot{\gamma}$}

\begin{figure*}[b]
    \includegraphics[scale=0.53, clip=true, trim=0mm 0mm 0mm 0mm]{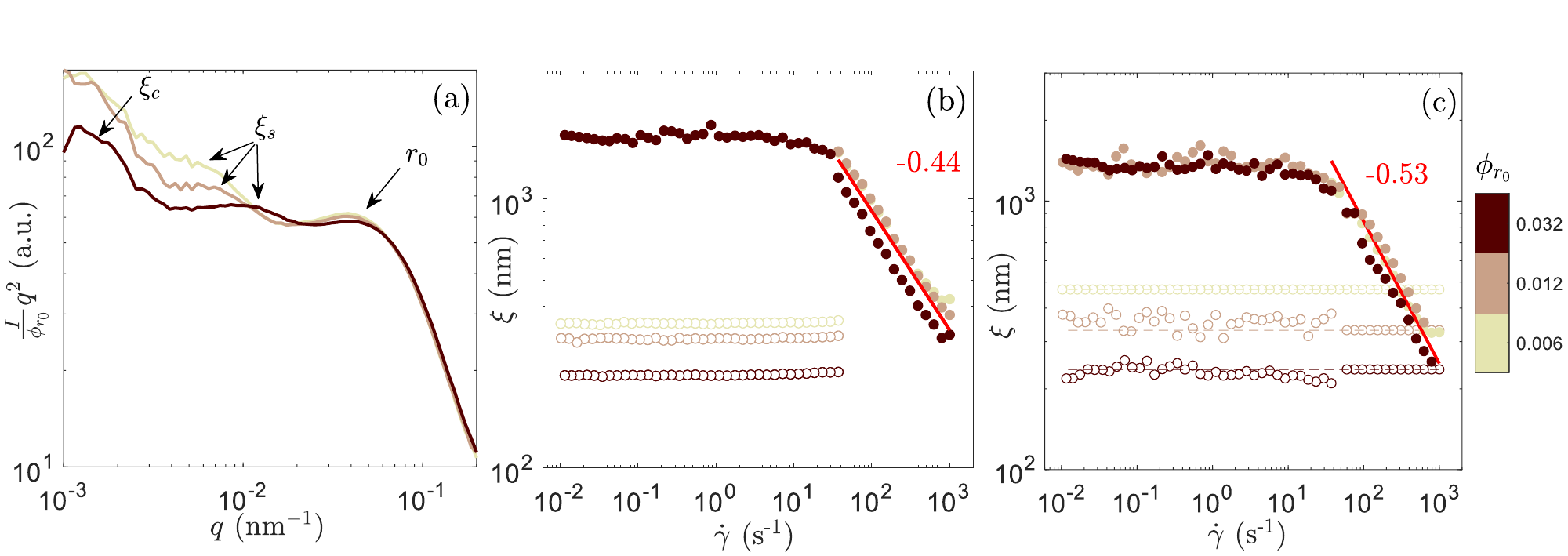}
    \centering
    \caption{(a) Rescaled scattering curve $I(q) \slash \phi{r_0}$ $vs$ $q$ measured under low shear rate at $\dot{\gamma}=0.02$~s$^{-1}$ for $\phi{r_0}$ = 0.006, 0.012 and 0.032. (b)-(c) Cluster size $\xi_c$ (solid circles) and intermediary length $\xi_s$ (empty circles) as function of the shear $\dot{\gamma}$ determined from the model-free analysis (b) and the modified two-level Beaucage model (c). Red lines are the best power law fit of $\xi_c$ $vs$ $\dot{\gamma}$ for $\dot{\gamma} > 50$ s$^{-1}$.}
    \label{fig:Xi_s}
\end{figure*}

The dependence of the length scales $\xi_s$ and $\xi_c$ with $\phi_{r_0}$ and $\dot{\gamma}$ was investigated using (i) a model free analysis (Eq.~\ref{eq:baeza}) and (ii) a two-level modified Beaucage model (Eq.~\ref{eq:Beaucage1}). $\xi_s$ and $\xi_c$ are displayed as function of $\dot{\gamma}$ on Fig.~\ref{fig:Xi_s}b-c.
Both approaches show that for $\dot{\gamma} > 50$ s$^{-1}$, $\xi_c$ increases with $\dot{\gamma}$ as $\xi_c \propto \dot{\gamma}^{-m}$ with $m \approx 0.5$, highlighting the robustness of this result. Below $\dot{\gamma} = 50$ s$^{-1}$, $\xi_c$ is found to be constant ($1 < \xi_c < 2$ \textmu m) at $\phi_{r_0}=0.032$. For lower volume fraction we believe the cluster size increases beyond the range of length scales that can be probed by the SAXS set-up. In the microstructural analysis developed in section~\ref{s:results} for the Fig.~\ref{fig:hydro}, we have extrapolated the values of $\xi_c$  using $\xi_c \propto \dot{\gamma}^{-m}$ up to the limit when $\sigma_h>\sigma$. 

In Fig.~\ref{fig:Xi_s}b-c, $\xi_s$ is found to be independent of the shear rate and slightly diminishes as $\phi_{r_0}$ increases. The dependence of $\xi_s$ with $\phi_{r_0}$ is also illustrated in Fig.~\ref{fig:Xi_s}a.

\clearpage 

%

\end{document}